\documentclass[11pt,letterpaper]{article}
\usepackage[top=0.85in,left=2.75in,footskip=0.75in]{geometry}

\usepackage{changepage}

\usepackage[utf8]{inputenc}

\usepackage{textcomp,marvosym}

\usepackage{fixltx2e}

\usepackage{amsmath,amssymb}

\usepackage{cite}

\usepackage{nameref,hyperref}

\usepackage{microtype}
\DisableLigatures[f]{encoding = *, family = * }

\usepackage{rotating}

\usepackage{multirow,booktabs}
\usepackage{amsmath,amssymb,latexsym,MnSymbol}
\usepackage{hyperref}

\raggedright
\usepackage[aboveskip=1pt,labelfont=bf,labelsep=period,justification=raggedright,singlelinecheck=off]{caption}

\makeatletter
\renewcommand{\@biblabel}[1]{\quad#1.}
\makeatother

\date{}

\usepackage{lastpage,fancyhdr,graphicx}
\usepackage{epstopdf}
\fancyheadoffset[L]{2.25in}
\fancyfootoffset[L]{2.25in}
\begin{document}
\vspace*{0.35in}

\begin{flushleft}
{\Large
\textbf\newline{The advantage of playing home in NBA: microscopic, team-specific and evolving features}
}
\newline
\\
Haroldo V. Ribeiro~\textsuperscript{1,*}, 
Satyam Mukherjee~\textsuperscript{2,3},
Xiao Han T. Zeng~\textsuperscript{4},
\\
\bigskip
\bf{1} Departamento de F\'isica, Universidade Estadual de Maring\'a, Maring\'a, PR 87020-900, Brazil
\\
\bf{2} Northwestern Institute on Complex Systems, Northwestern University, Evanston, IL 60208, USA
\\
\bf{3} Indian Institute of Management, Udaipur, India
\\
\bf{4} Groupon, Inc. Chicago, IL 60654, USA
\bigskip

%
%





* \url{hvr@dfi.uem.br}

\end{flushleft}
\section*{Abstract}
The idea that the success rate of a team increases when playing home is broadly accepted and documented for a wide variety of sports. Investigations on the so-called ``home advantage phenomenon'' date back to the 70's and every since has attracted the attention of scholars and sport enthusiasts. These studies have been mainly focused on identifying the phenomenon and trying to correlate it with external factors such as crowd noise and referee bias. Much less is known about the effects of home advantage in the ``microscopic'' dynamics of the game (within the game) or possible team-specific and evolving features of this phenomenon. Here we present a detailed study of these previous features in the National Basketball Association (NBA). By analyzing play-by-play events of more than sixteen thousand games that span thirteen NBA seasons, we have found that home advantage affects the microscopic dynamics of the game by increasing the scoring rates and decreasing the time intervals between scores of teams playing home. We verified that these two features are different among the NBA teams, for instance, the scoring rate of the Cleveland Cavaliers team is increased $\approx0.16$ points per minute (on average the seasons 2004-05 to 2013-14) when playing home, whereas for the New Jersey Nets (now the Brooklyn Nets) this rate increases in only $\approx0.04$ points per minute. We further observed that these microscopic features have evolved over time in a non-trivial manner when analyzing the results team-by-team. However, after averaging over all teams some regularities emerge; in particular, we noticed that the average differences in the scoring rates and in the characteristic times (related to the time intervals between scores) have slightly decreased over time, suggesting a weakening of the phenomenon. This study thus adds evidence of the home advantage phenomenon and contributes to a deeper understanding of this effect over the course of games.

  

\section*{Introduction}
Competitive events among agents or groups are ubiquitous in nature and society. Understanding these competitive processes is a natural academic goal that finds important applications in economics, politics, and sports. In particular, sports are considered a natural laboratory for testing hypotheses and studying competitions~\cite{Kahn,BenNaim}, with the tremendous advantage of offering more and more datasets that not only provide results or summaries of massive amounts of games, but also enable a complete recap of the within-game events of entire seasons of sport leagues. This unprecedented amount of data enabled scholars to probe patterns of such competitive events to a degree not before possible, answering many academic questions as well as elucidating sport folklores. Examples of such investigations include random walks or diffusive interpretations of the scoring process~\cite{BenNaim2,BenNaim3,Sire,Sire2,Heuer,Ribeiro,Gabel,Ribeiro2,Ribeiro3,Clauset}, discussions about the efficiency of sport competitions~\cite{BenNaim3,Ribeiro,BenNaim4,Erculj,DeSaa,DeSaa2,DeSaa3}, analysis of player and team performance via networks tools~\cite{Duch,Radicchi,Fewell,Mukherjee,Mukherjee2} and tracking data~\cite{Sampaio}, performance evolution in the Olympic Games~\cite{Radicchi2}, the role of coaching experience in the effective use of timeouts~\cite{Saavedra}, reciprocity in passing patterns~\cite{Willer}, cooperative play~\cite{Uhlmann}, Matthew effect in the longevity of careers in professional sport~\cite{Petersen,Perc}, and the hot-hand phenomenon~\cite{Ribeiro2,Yaari,Bock,Yaari2,Csapo}.

Success in sport competitions is not only a fan demand, but also a long-standing business involving billions of dollars, management, finance, and marketing policies~\cite{Rosner}. Thus, the identification of key factors that have a systematic influence on the success rate of teams and athletes goes beyond a theoretical question and may attract the interest of teams, coaches and players as well. 

One of the consistent factors that are likely to affect the success rate in sport competitions is the game location. Despite some controversial findings, the idea that the success rate of a team (or a player) increases when playing home is widely accepted and documented for several sports. Starting with the seminal work of Schwartz and Stephen~\cite{Schwartz} in 1977, the ``home advantage phenomenon'' has motivated several investigations ever since~\cite{Courneya,Nevill2,Nevill1}. A non-exhaustive list of sports where this phenomenon has been found include soccer~\cite{Pollard,Nevill,Pollard3,Page,Monks,Riedl,Staufenbiel}, baseball~\cite{Bray,Levernier}, ice hockey~\cite{McGuire,Agnew}, roller-hockey~\cite{Gomezrh}, basketball~\cite{Jones,PollardB}, rugby~\cite{Thomas}, Australian football~\cite{Clarkeaf}, water polo~\cite{Prietowp}, volleyball~\cite{Marcelino}, handball~\cite{Pollardhb,Oliveira,Smiatek}, and cricket~\cite{Morley}. This effect has also been observed in individual competitions of tennis~\cite{Koning,Nevill4}, golf~\cite{Nevill4}, Winter Olympics sports~\cite{Balmer}, Summer Olympics sports~\cite{Balmerso}, and several other individual sports~\cite{Jones2}. Jamieson~\cite{Jamieson} reported an interesting meta-analysis on several sports (including some of the above-cited), where it was found that the overall home winning percentage is about $60$\%, and moderator factors such as time era (matches prior to the 50's are more affected by this phenomenon than more recent ones) and sport (home advantage is more intense for soccer than several sports) were also identified. Researchers have also tried to assign causes related to the home advantage such as crowd noise~\cite{Nevill3,Myers}, audience hostility~\cite{Anders}, away-team travels~\cite{Goumas}, tactics used by teams and coaches~\cite{Gomezta,Pollardta}, familiarity with the local playing facility~\cite{Pollard2} and referee bias~\cite{Johnston} as well as pointed differences between teams from capital and inner cities~\cite{Gomes}. 

Despite this considerable interest, much less is known about the effects of home advantage in the ``microscopic'' dynamics of games, that is, the changes this phenomenon causes within game events. Team-specific and evolving features of this phenomenon are other questions that are also rarely tackled in the previous pages. Exceptions include the works of Pollard and Pollard, which studied the evolution of the winning percentage at home for team sports~\cite{Pollard5} and regional variations in this percentage~\cite{Pollard6}. The former aspect along with a comparison between men and women soccer leagues was also discussed by Pollard and G\'omez~\cite{Pollard4}. There are also evidence that home advantage is more intense at the beginning of basketball matches~\cite{Jones}, time dependent for handball~\cite{Oliveira}, and that the scoring processes is highly dynamic~\cite{DeSaa2} for basketball. All these features raise several questions on the microscopic effects of the home advantage, and also on how these features may possibly differ among teams and time era.

In this article, we present a detailed study of the effects of home advantage in microscopic dynamics of more than sixteen thousand games spanning thirteen seasons of National Basketball Association (NBA). By analyzing the play-by-play events of the games, we find that the scoring rates increase when teams play home, whereas the time intervals between scores decrease. We have further observed that these features vary across teams, seasons, and game time (quarters). The overall average differences in the scoring rates and in the characteristic time intervals have slightly decreased over time, suggesting that home advantage has become weaker in the NBA. We also report a rank of the NBA teams according to the intensity of the home advantage in these two microscopic features of the game.

\section*{Methods}
\subsection*{Data presentation}
We have accessed data from the official web portal of ESPN, under the NBA section: \url{http://espn.go.com/nba/}. By browsing under the URL \url{http://espn.go.com/nba/schedule/}, we initially obtained all game identification between the years of 2001 and 2014. This game identification leads to a web page that contains information about the game, including game place and the play-by-play recap of game events (points, missing points, rebound, etc), see \url{http://espn.go.com/nba/playbyplay?gameId=400489378} for an example of such pages for the game Detroit Pistols versus New York Knicks (playing home) in Jan 7, 2014. We thus downloaded all available game information pages, grouping the games according to the NBA season and removing special games such as NBA All-Star Games and matches involving foreign teams (Olympiacos, FC Barcelona, CSKA Moscow, etc). From these pages, we extracted the team names, game place, match date, and the score evolution $S(t)$ of each team as a function of the game time $t$. At this stage, we further removed games for which no play-by-play events were available and also games in which the score evolution was not a monotonically increasing function of time $t$. These lead us to 16,133 games covering 13 NBA seasons (from the 2001--02 to the 2013--14 season). A small random sample of these data was further manually compared with the play-by-play events from the NBA official web page, and a perfect agreement was found for the score events. As these data are subject to updates, a snapshot has been provided as \nameref{S1_dataset}. 

\section*{Results and Discussion}

\begin{figure}[!ht]
\begin{adjustwidth}{-2.25in}{0in}
\begin{center}
\includegraphics[scale=0.48]{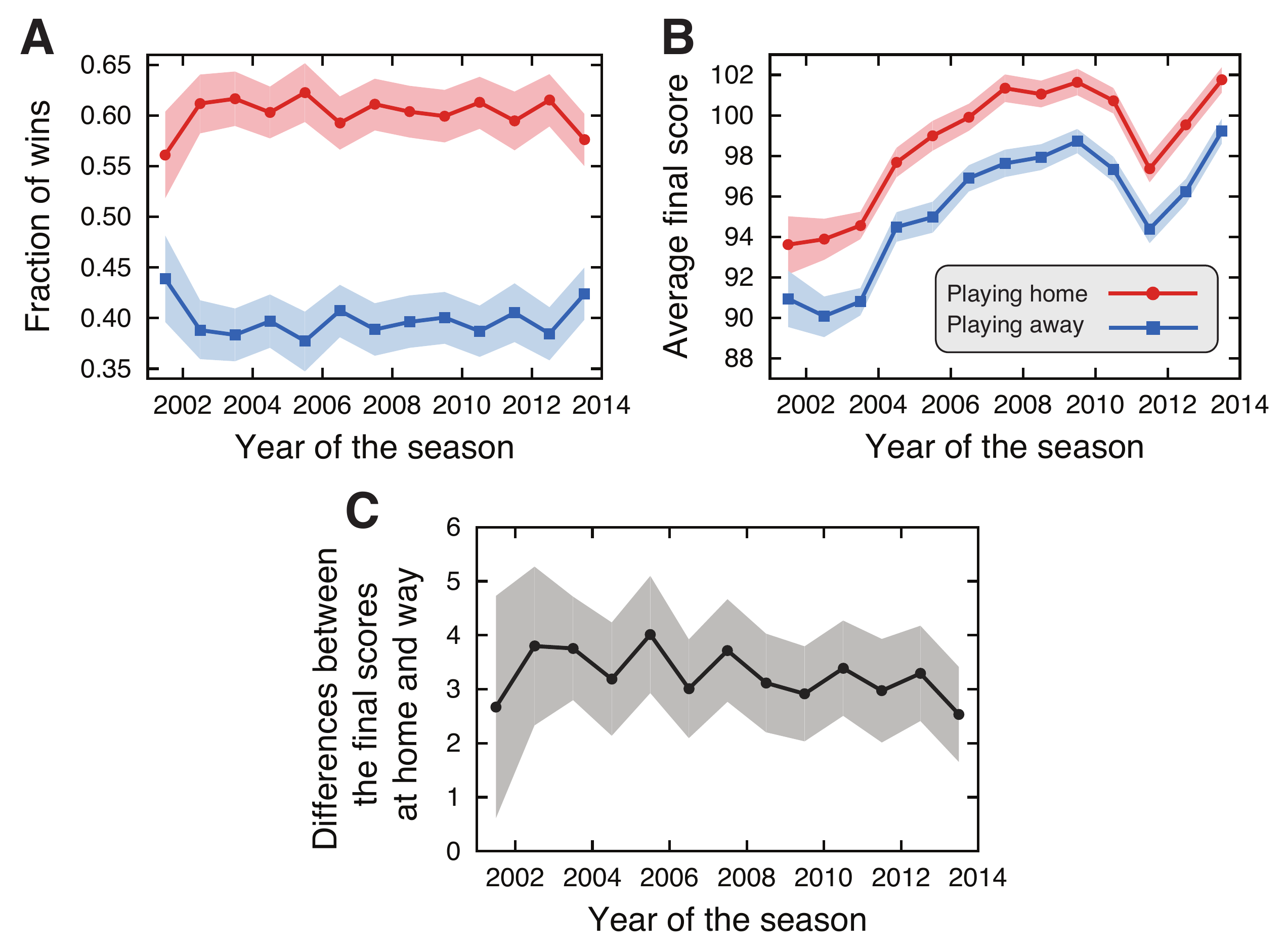}
\end{center}
\caption{\textbf{Macroscopic manifestation of the home advantage in NBA.} (A) Average fraction of wins when playing home (red circles) and away (blue squares) along the thirteen NBA seasons studied here. (B) Average final score of the teams when playing home (red circles) and away (blue squares). (C) Evolution of the differences between the final scores and at home and away along the NBA seasons. In all plots, the shaded areas stand for 95\% bootstrap confidence intervals.}
\label{fig:1}
\end{adjustwidth}
\end{figure}

We start by investigating the holistic idea of home advantage, that is, we ask whether the teams playing home have a large fraction of wins than those playing away. We calculate this fraction for each NBA season available in our dataset and the results are depicted in Fig~\ref{fig:1}A. We observe that teams playing home wins about 60\% of the matches, 10\% more than would be expected by chance. Similar values were reported for several basketball leagues~\cite{Jones,PollardB} and also in the meta analysis of Jamieson~\cite{Jamieson}. It is also worth noting that this fraction is almost constant over the years. Another manner of quantifying the macroscopic effect of home advantage is by evaluating the average final score of teams playing home and away. Fig~\ref{fig:1}B shows these quantities for each NBA season, where we (naturally) observe that teams playing home have greater final scores than those playing away. Interestingly, we notice that the average final scores show an evolving behavior: between seasons 2001--02 and 2009--10 these averages have increased, followed by a sharp decrease over the next two seasons (2010--11 and 2011--12) and again by an increasing behavior for the last two seasons, reaching almost the same values as in 2011--12. We have further investigated the difference between the final scores at home and away, as shown in Fig~\ref{fig:1}C. We observe that teams playing home score $3.3\pm0.1$ points (average of all seasons) more than those playing away; also, this difference seems to present decreasing trend. 

In order to quantify this trend, a linear regression model was adjusted to these data, yielding a weak decreasing trend of $0.08\pm0.03$ points per year. It is worth noting that this regression does not account for team ability, which may introduce some bias in this evolving behavior. As discussed by Pollard and G\'omez~\cite{Pollard4}, more balanced leagues are more likely to be strongly affected by home advantage. Thus, this evolving trend could also be related to changes in the competitiveness of NBA seasons. However, studies have suggested that the competitive balance in NBA is stable over time and close to its average~\cite{DeSaa,DeSaa2} and that the scoring process is well described by a nearly unbiased random walk~\cite{Gabel}, suggesting that NBA teams are quite balanced. Furthermore, a decrease in the intensity of home advantage over time was also observed by Jamieson~\cite{Jamieson} for several sports. Therefore, despite the lack of a more precise approach for quantifying the evolving trend of home advantage (which is interesting but out of the scope of this article), our results are in agreement with recent findings on the subject and may be just slightly affected by the differences in ability among teams. Similar discussions apply to forthcoming analysis on evolving trends of other aspects of home advantage.

\begin{figure}[!ht]
\begin{adjustwidth}{-2.25in}{0in}
\begin{center}
\includegraphics[scale=0.53]{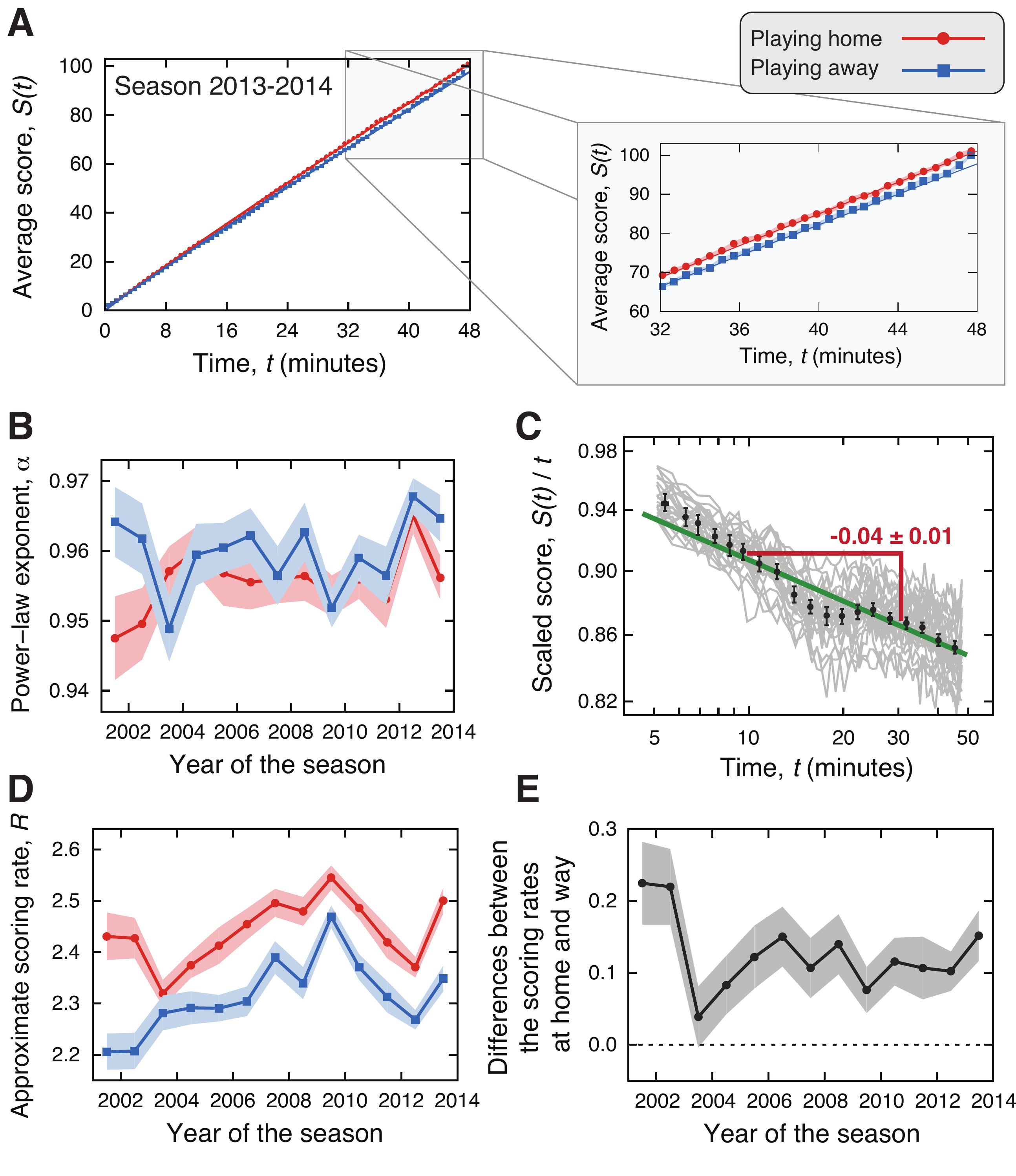}
\end{center}
\caption{\textbf{Evidence for home advantage in the score evolution.} (A) Average score $S(t)$ as a function of the game time $t$ when playing home (red circles) and away (blue squares). These averages were calculated for the NBA season 2013--14 (see \hyperref[S1_Fig]{S1}~Fig for all seasons). The last 16 minutes of the games are highlighted. The continuous lines (red for home and blue for away) represent the adjusted power-law models [$S(t) = R\, t^{\alpha}$]. Notice that the difference between the average scores at home and away increases over time. (B) Evolution of the power-law exponent $\alpha$ over the NBA seasons. We observe practically no difference between playing home and away; however, the values of $\alpha$ are all smaller than one, indicating that the score evolution is slightly sub-linear. (C) The gray curves show the average score $S(t)$ divided by $t$ as a function of the game time $t$ calculated for every NBA season and grouping the matches by field. The black dots are window average values over all curves and the error bars stand for 95\% confidence intervals. The green line is a power-law fit to average tendency whose slope (power-law exponent) is $0.04\pm 0.01$. (D) Evolution of the approximate scoring rates $R$ over the NBA seasons. The teams playing home display significantly larger rates (average over all seasons of $2.44\pm0.02$ points per minute) than when playing away ($2.31\pm0.02$ points per minute). (E) Evolution of the differences between the scoring rates at home and away along the NBA seasons. Notice that these values display a decreasing tendency over the years. The shaded areas in the plots stand for 95\% bootstrap confidence intervals.}
\label{fig:2}
\end{adjustwidth}
\end{figure}

The previous analysis has thus confirmed the existence of home advantage in the NBA games; however, it provides no clues on how this advantage emerge from the microscopic dynamics of the games. In order to investigate such aspects, we calculate the average score $S(t)$ as a function of the game time $t$, after grouping the matches by seasons and field (home or away). Fig~\ref{fig:2}A shows an example of the behavior of $S(t)$ for teams playing home and away in the season 2013--14. We observe that $S(t)$ increases faster for teams playing home than those playing away and that a statistically significant difference appears around $t \approx 16$~minutes. Similar behaviors are observed in all seasons (see \hyperref[S1_Fig]{S1}~Fig). 

\clearpage
Another intriguing feature of the behavior of $S(t)$ is that it appears to increase as a nonlinear function of time, which is visible from its slight concave shape. To verify this nonlinear behavior, we have adjusted a linear function to the relationship $\log S(t)$ versus $\log t$; in this case, a unitary linear coefficient indicates that $S(t)$ increases linearly in time and deviations from the unitary value point out for a power-law behavior in $S(t)$, that is, 
\begin{equation}\label{eq:scoreevo}
S(t)= R\, t^\alpha\,,
\end{equation}
with $\alpha$ being the power-law exponent (or the linear coefficient in the log-log relationship) and $R$ a multiplicative constant (or the intercept in the log-log relationship). Fig~\ref{fig:2}B shows the values of $\alpha$ for playing home and away in each NBA season. We observe that these values are practically identical regarding playing home or away and that they can be well approximated by a constant plateau; however, we do observe that the values of $\alpha$ are all smaller than one, indicating that the scores increase (slightly) sub-linearly in time. Fig~\ref{fig:2}C shows the average score $S(t)$ divided by $t$ as a function of time for each season as well as for playing home and away. If the average score was linear in time, these curves would be approximated by horizontal lines in these log-log plots; instead, we observe a decreasing behavior. Furthermore, by fitting a power-law function to the average behavior of $S(t)/t$ versus $t$, we find that the power-law exponent is $0.04\pm0.01$, a value that is consistent with the values of $\alpha$ reported in Fig~\ref{fig:2}B, that is, for $S(t)\sim t^{\alpha}$, we expect $S(t)/t\sim t^{\alpha-1}$. 

The sub-linear behavior of the scores versus time indicates the scoring rates are not constant over time, as was also observed by Gabel and Redner~\cite{Gabel}; actually, the values of $\alpha<1$ indicate that the scoring rates decrease with the passing of time --- a fact probably related to the physical wear of the athletes along the game. However, the values of $\alpha$ are not very different from one, and we may consider that the value of $R$ represents an approximate the scoring rate. Fig~\ref{fig:2}D shows the values of $R$ estimated for teams playing home and away for every NBA season. We notice that teams playing home have a statistically significantly larger scoring rate than those playing away. By averaging over all seasons in our dataset, we estimate that teams score at $R=2.44 \pm 0.02$ points per minute at home, whereas $R=2.31 \pm 0.02$ points per minute is the average scoring rate in away matches. We further calculate the difference between the values of $R$ at home and away, as shown in Fig~\ref{fig:2}E. Despite the sharp change occurred in the 2002--03 season and similarly to the results of Fig~\ref{fig:1}, the differences between the values of $R$ seems to decrease over time. A linear regression model adjusted to these data indicates that difference in the scoring rates is diminishing at a slight pace of $0.004 \pm 0.002$ points per minute per year.%

\begin{figure}[!ht]
\begin{adjustwidth}{-2.25in}{0in}
\begin{center}
\includegraphics[scale=0.288]{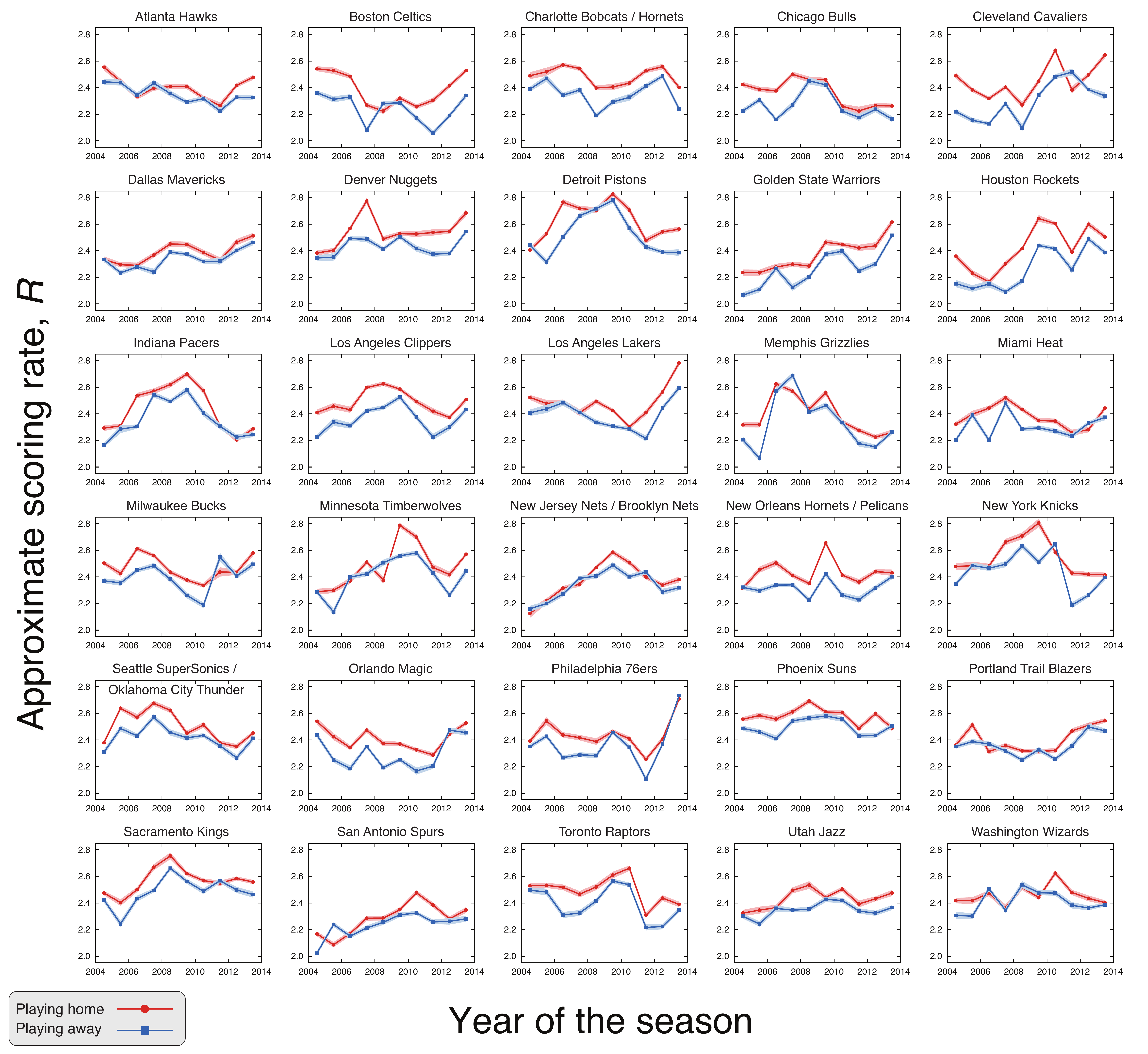}
\end{center}
\caption{\textbf{Evolution of the scoring rate when playing home and away for each NBA team.} The panels show the approximate scoring rates $R$ when playing home (red circles) and away (blue squares) for every team and season from 2004--05 to 2013--14, period in which the teams were the same. The shaded areas are 95\% bootstrap confidence intervals. Notice that the scoring rates are systematically larger when the team plays home; however, we do observe some inversions and that values of $R$ vary among teams and seasons.}
\label{fig:3}
\end{adjustwidth}
\end{figure}

\begin{figure}[!ht]
\begin{adjustwidth}{-2.25in}{0in}
\begin{center}
\includegraphics[scale=0.45]{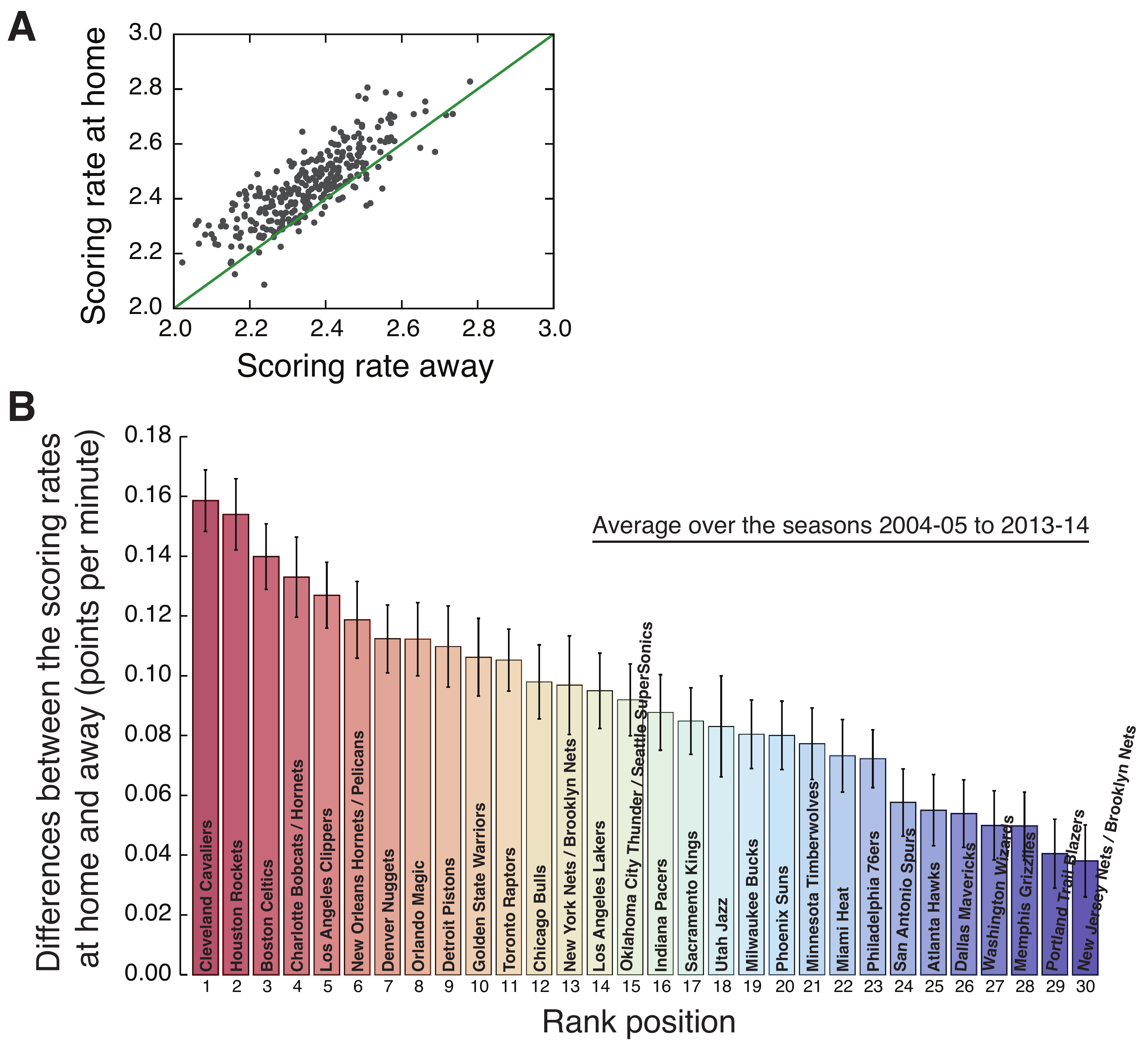}
\end{center}
\caption{\textbf{Ranking NBA teams according to the difference between the scoring rate at home and away.} (A) Scoring rate at home versus scoring rate away (that is, the values of $R$). The dots represent the scoring rates (at home and away) for every team and NBA season, and the green line is a linear function (with a unitary linear coefficient and a zero intercept). Notice that there are only a few cases in which the scoring rate is larger when playing away than when playing home. (B) Average of the difference between the scoring rates at home and away for each NBA team (in descending order). These averages were calculated over the seasons 2004--05 to 2013--14 (during this period the teams were the same; see and \hyperref[S6_Fig]{S6}~Fig for all scoring rates) and the error bars are 95\% bootstrap confidence intervals.}
\label{fig:4}
\end{adjustwidth}
\end{figure}

The approximate scoring rate $R$ previously described is an average over all teams. This begs the intriguing question whether the values of $R$ differs from team to team. To address this question, we estimate the values of $R$ at home and away for each NBA team between the seasons 2004--05 and 2013--14. In this period, the number of teams was fixed as thirty (current number) and the same teams competed in the league. The only change in the list of teams occurred at the end of the 2007--08 season, when the team Seattle SuperSonics was relocated to Oklahoma City and now plays as the Oklahoma City Thunder. We have considered this event only as a name change and assumed the matches before and after the 2007--08 are from the same team. In order to estimate the values of $R$ for each team and focus only on its evolution, we have fixed the values of $\alpha$ to its overall average value when fitting the relationships between $\log S(t)$ and $t$ (Eq.~\ref{eq:scoreevo}). Fig~\ref{fig:3} shows the values of $R$ for each team and season. Despite a few exceptions, we observe that the scoring rates are systematically larger when playing home than away for all teams. In Fig~\ref{fig:4}A, we plot the values of $R$ at home against the values of the $R$ estimated for away matches for each team and season, where it is further evident that the occurrence of teams which larger scoring rates in away matches in a NBA season is very rare (around 1\% of the teams by season). 

We further observe that despite the complicated evolving behavior of $R$ reported in Fig~\ref{fig:3} for each team, one can note that teams such as New Jersey Nets (now the Brooklyn Nets) and Portland Trail Blazers have small differences between the scoring rates at home and away when compared with other teams such as Cleveland Cavaliers and Houston Rockets. To quantify these differences, we have estimated the scoring rates at home and away from the evolution scores $S(t)$ averaged over the seasons 2004--05 to 2013--14 for each NBA team (see \hyperref[S2_Fig]{S2}~Fig). Fig~\ref{fig:4}B shows a rank of the teams according to the difference between the scoring rates at home and away and confirms the existence of statistically significant differences among them. For instance, Cleveland and Houston have the largest differences and score about $0.16$ point per minute more when playing home, whereas New Jersey and Portland have the smallest differences (around $0.04$ point per minute more in home matches). 

\begin{figure}[!hb]
\begin{adjustwidth}{-2.25in}{0in}
\begin{center}
\includegraphics[scale=0.42]{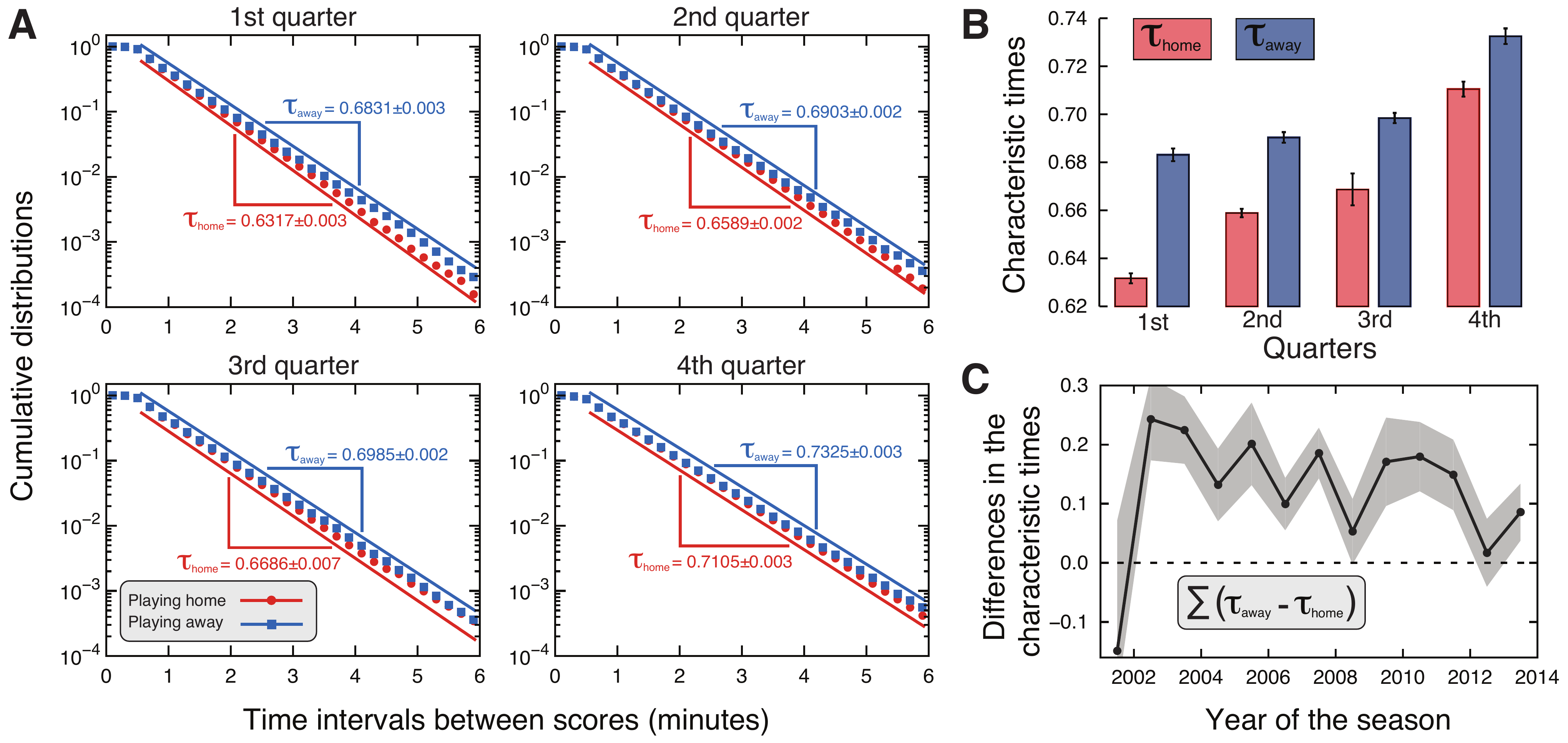}
\end{center}
\caption{{\bf Evidence for home advantage in the time intervals between scores.} (A) Cumulative distributions of the time intervals between stores when the teams play home (red dots) and away (blue squares). The panels show the distributions for the four quarters (period of 12 minutes in which the games are played). Here we have aggregated data from all seasons (see \hyperref[S3_Fig]{S3}, \hyperref[S4_Fig]{S4}, \hyperref[S5_Fig]{S5} and \hyperref[S6_Fig]{S6}~Figs for individual results). All distributions are well approximated by exponential distributions, that is, $P(\Delta t)\sim e^{-\Delta t/\tau}$, where $\tau=\tau_{\;\text{home}}$ is the characteristic time interval when playing home and $\tau=\tau_{\;\text{away}}$ is the analogous when playing away. Notice that the plots are in log-lin scale and thus the exponential decay is linearized. The values of $\tau_{\;\text{home}}$ and $\tau_{\;\text{away}}$ were estimated via maximum likelihood method and are shown in the plots. The straight lines are guides for the eyes indicating the adjusted behavior of $P(\Delta t)$. We observe that these distributions decay faster for teams playing home than playing away. (B) Bar plots of the characteristic times $\tau_{\;\text{home}}$ and $\tau_{\;\text{away}}$ for each quarter. The error bars stand for 95\% bootstrap confidence intervals. The characteristic times are systematically smaller when the teams play home than when playing way; we further observe that the difference $\tau_{\;\text{away}} - \tau_{\;\text{home}}$ decreases with the passing of the quarters. (C) Evolution of the sum of the differences between the characteristic times at home and away [$\sum (\tau_{\;\text{away}} - \tau_{\;\text{home}})$,  over all quarters] along the NBA seasons. The shaded areas stand for 95\% bootstrap confidence intervals.}
\label{fig:5}
\end{adjustwidth}
\end{figure}

We now focus on quantifying the role of playing home in another microscopic game feature: the time intervals between scores. In this context, it is natural to imagine that teams playing home may display a faster rhythm, perhaps driven by the home team crowd~\cite{Nevill,Myers,Anders}. In order to investigate this possibility, we have estimated the probability distributions of the time intervals between scores in each quarter of the game. Fig~\ref{fig:5}A shows these distributions when aggregating data from all seasons and grouping home and away matches. We note that these empirical distributions are well described by exponential distributions, that is,
\begin{equation}
P(\Delta t) = (1/\tau) \exp(-\Delta t/\tau)\,,
\end{equation}
where $P(t)$ is the probability of finding a time interval between scores equal to $\Delta t$ and $\tau$ is the characteristic time interval (the only distribution parameter). Similar exponential distributions were reported by Gabel and Redner~\cite{Gabel} when considering all quarters together. In addition, we observe that the exponential decays of these distributions are faster for the time intervals occurring in home matches than in away matches. To quantify this difference, we estimate (via maximum likelihood method) the values of the characteristic time interval $\tau$ when playing home ($\tau=\tau_{\text{home}}$) and away ($\tau=\tau_{\text{away}}$) for each quarter. The values of these parameters are shown in Fig~\ref{fig:5}A as well as are represented in a bar plot in Fig~\ref{fig:5}B. We find that the characteristic time interval between scores is statistically significant smaller in home matches than in away matches in all quarters. It is worth noting that the values of $\tau_{\text{home}}$ and $\tau_{\text{away}}$ are also the average value of the time intervals $\Delta t$; thus, home teams actually have a faster scoring pace. We observe that both the values of $\tau_{\text{home}}$ and $\tau_{\text{away}}$ increase along the quarters of the game; again, a fact that is likely caused by the fatigue process of the players. However, we do notice that the largest gap between $\tau_{\text{home}}$ and $\tau_{\text{away}}$ occurs in the first quarter and that this gap is gradually reduced as the game progresses. This result is in agreement with findings of Jones~\cite{Jones} and suggests that this aspect of the home advantage is stronger at the beginning of the matches, which could be caused by the usual intense reception of the home team by its fans or also by the initial unfamiliarity of the guest team with the arena and its audience. 

We have also evaluated the probability distributions of the time intervals between scores after grouping our data by NBA season. We show these distribution for each game quarter in \hyperref[S3_Fig]{S3}, \hyperref[S4_Fig]{S4}, \hyperref[S5_Fig]{S5} and \hyperref[S6_Fig]{S6}~Figs, where similar exponential behaviors also emerge. Once again, we estimate the parameters $\tau_{\text{home}}$ and $\tau_{\text{away}}$ via maximum likelihood method for each quarter and season. Fig~\ref{fig:5}C shows the evolution of the sum of the differences between $\tau_{\text{away}}$ and $\tau_{\text{home}}$ over the quarters. Despite the some fluctuations and similarly to our previous results (Figs~\ref{fig:1}C and~\ref{fig:2}E), the differences in the characteristic time intervals between scores appear to decrease over time. A linear regression on this trend (after discarding the season 2001--02) indicates a decreasing tendency of $0.012 \pm 0.005$ minutes per year.

\begin{figure}[!ht]
\begin{adjustwidth}{-2.25in}{0in}
\begin{center}
\includegraphics[scale=0.28]{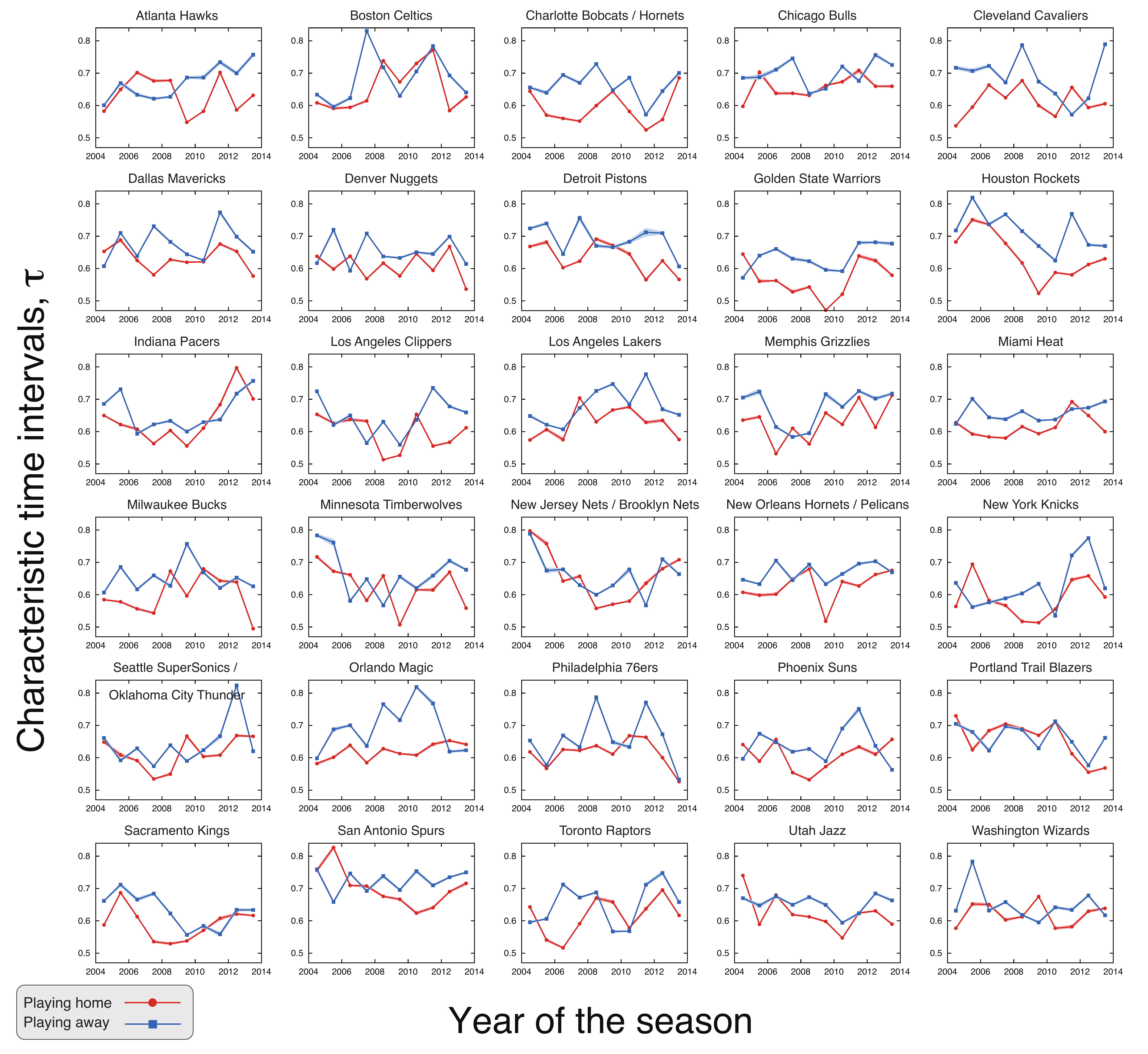}
\end{center}
\caption{\textbf{Evolution of the characteristic time intervals when playing home and away for each NBA team.} The panels show the characteristic time intervals between scores when playing home ($\tau_{\;\text{home}}$, red circles) and away ($\tau_{\;\text{away}}$, blue squares) for every team and season from 2004--05 to 2013--14, period in which the teams were the same. Here we have aggregated data from all quarters and estimated the characteristic times via maximum likelihood method. The shaded areas are 95\% bootstrap confidence intervals. Notice that the characteristic times are systematically larger when the team plays away; still, we note some inversions and that the values vary among teams and seasons.}
\label{fig:6}
\end{adjustwidth}
\end{figure}

\begin{figure}[!ht]
\begin{adjustwidth}{-2.25in}{0in}
\begin{center}
\includegraphics[scale=0.45]{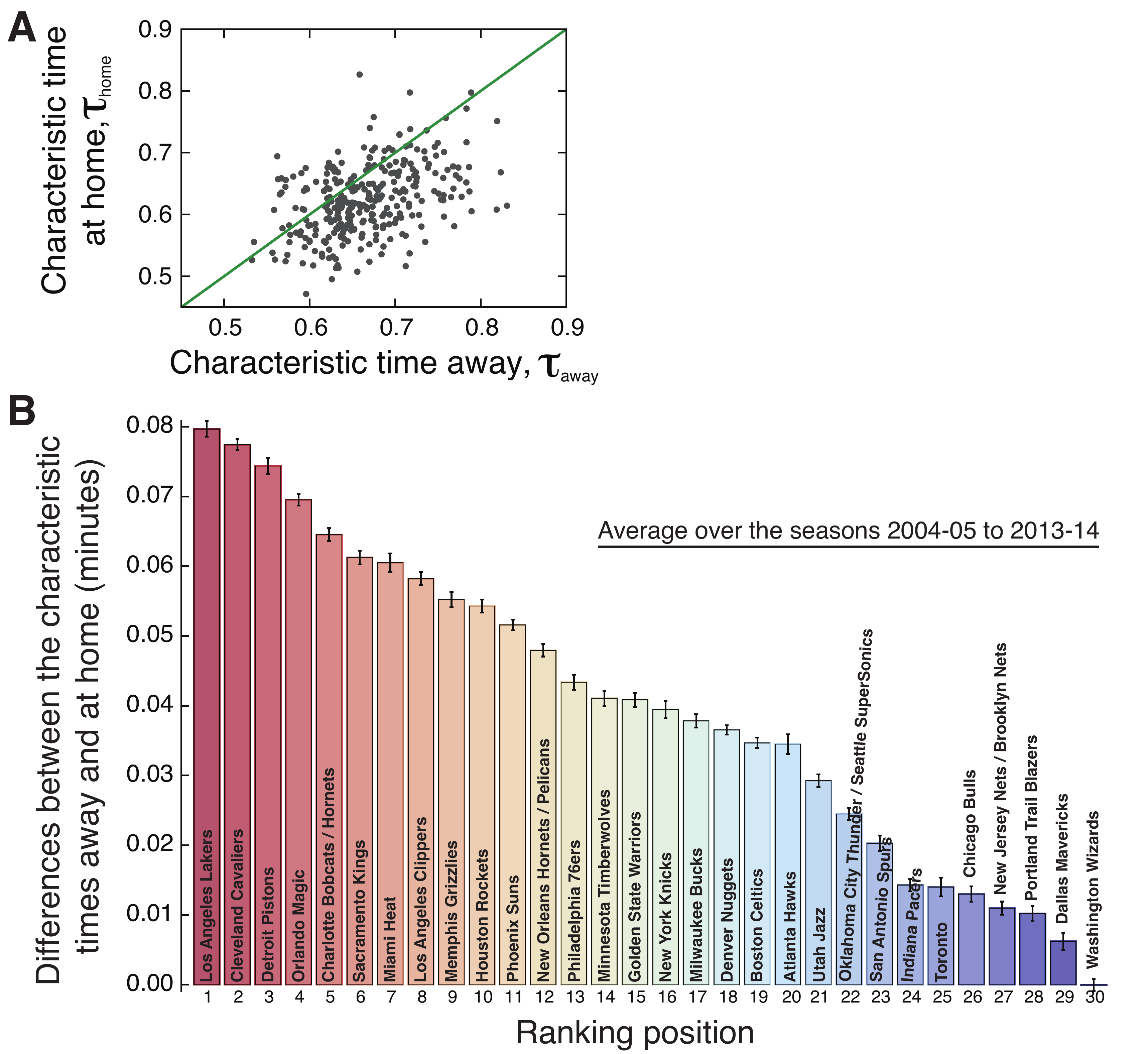}
\end{center}
\caption{\textbf{Ranking NBA teams according to the difference between the characteristic time intervals at home and away.} (A) Characteristic time intervals between scores at home ($\tau_{\;\text{home}}$) versus away ($\tau_{\;\text{away}}$). The dots represent the values of the characteristic times (at home and away) for every team and season from 2004--05 to 2013--14 (during this period the teams were the same), and the green line is a linear function ($\tau_{\;\text{home}} = \tau_{\;\text{away}}$). Here we have aggregated data from all quarters and estimated the characteristic times via maximum likelihood method. Notice that there are only a few cases in which the characteristic time is larger when playing home than when playing away. (B) Average of the difference between $\tau_{\;\text{away}}$ and $\tau_{\;\text{home}}$ for each NBA team (in descending order). These averages were calculated over the seasons 2004--05 to 2013--14 (see \hyperref[S7_Fig]{S7}~Fig for the cumulative distributions) and the error bars are 95\% bootstrap confidence intervals.}
\label{fig:7}
\end{adjustwidth}
\end{figure}

Finally, we turn our attention to possible team-specific features related to the time intervals between scores. To do so, we proceed as it was did for the scoring rates, that is, we have estimated the values of the characteristic times ($\tau_{\text{home}}$ and $\tau_{\text{away}}$) for each of the thirty teams that have competed in the league during the seasons 2004--05 and 2013--14. Different from the previous analysis, we have now aggregated data from all quarters for obtaining a more reliable estimate of the time intervals distribution and its parameter. Fig~\ref{fig:6} shows the values of the characteristic times for playing home and away (estimated via maximum likelihood method) for each team and season. We observe that values of $\tau$ are (despite some inversions) systematically larger when the teams play away from home, that is, the average time interval between scores is larger in away matches. Fig~\ref{fig:7}A shows a scatter plot of the values of $\tau_{\text{home}}$ versus $\tau_{\text{away}}$ for every team and season, where we observe that the occurrence of teams with larger characteristic times at home in a season is very rare (around 2\% of the teams by season). Despite that, we notice that this relationship exhibits larger fluctuations when compared with Fig~\ref{fig:4}A. However, the linear regression $\tau_{\;\text{home}} = b +  \tau_{\;\text{away}}$ finds $b = -0.043\pm0.004$ ($p$-value $<10^{-16}$), indicating that the occurrence of $\tau_{\;\text{home}}<\tau_{\;\text{away}}$ cannot be explained by chance.

Similarly to the case of the scoring rates, the evolving behavior of the characteristic times reported in Fig~\ref{fig:6} is somehow a noisy one (visually larger than that reported in Fig~\ref{fig:3}). Part of this behavior could be associated with the intrinsic changes of the teams; however, there may also be some fluctuations related to the different number of events employed when estimating values of the characteristic times. Thus, just as it was did when ranking the teams according to the scoring rates, we have aggregated the time interval events of the teams over the seasons 2004--05 to 2013--14 for estimating the distributions $P(\Delta t)$ as well as the parameters $\tau_{\text{home}}$ and $\tau_{\text{away}}$ (see \hyperref[S7_Fig]{S7 Fig}). Fig~\ref{fig:7}B shows a rank of the teams according to the difference $\tau_{\text{away}}-\tau_{\text{home}}$, where we observe statistically significant difference among teams. This difference ranges from about 0.08 minutes for the Los Angeles Lakers and Cleveland to around zero for the Washington. It is worth noting that the ranking based on the scoring rates (Fig~\ref{fig:4}B) and on the characteristic times are not the same, that is, there are several disagreements between the two rankings. However, they are correlated as measured by the Spearman's rank correlation coefficient  ($\rho=0.50$, $p$-value~$=0.005$) or by the Kendall's one~\cite{Sheskin} ($\rho=0.34$, $p$-value~$=0.009$), suggesting that teams playing home not only have higher scores but also score at a faster rate.

\clearpage
\section*{Conclusions}
In this work, we presented a new view of the home advantage phenomenon by focusing on the microscopic features of NBA matches. Specifically, we asked about the role of playing home on the dynamics of the score events within NBA games. Firstly, we studied the time behavior of the scores along the games, where it was found that the average score increases slightly sub-linearly in time for home and away matches. Based on this behavior, we defined an approximate scoring rate and determined that teams score an average of $0.13$ points per minute more in home matches. We also verified that this number appears to be diminishing over the seasons at a slight pace, a behavior that also appears in other sports~\cite{Jamieson}. We further estimated the scoring rates at home and away for every team and season (from 2004--05 to 2013--14), where we observed that the difference between these rates changes from team to team. A ranking of teams according to the difference between the scoring rates at home and away was also presented. Next, we focused our attention on the time intervals between scores. The probability distribution of these times was found to be in good agreement with an exponential distribution, where the characteristic times for away matches are larger than the values for home matches. We noticed that this gap is gradually reduced over the game progress (that is, along the game quarters), which indicates that home advantage in NBA is mostly accumulated in the beginning of the matches, as was also discussed by Jones~\cite{Jones}. In addition, the difference in the characteristic times has decreased over the NBA seasons. Analogous to the scoring rates, the difference in the characteristic times at home and away is a team-specific feature, which enabled us to rank the teams according to this difference. Both the reduction in the difference between the scoring rates and in the characteristic time intervals suggest that the microscopic effect of the home advantage phenomenon is slowly becoming weaker. This change might be because teams are using better strategies to overcome disadvantages when playing away. 

Our thus work provides new clues about the role of playing home and away in sport competitions by showing how two microscopic features of NBA games are affected. We further believe that our approach could be useful for a better understanding of the universality of home advantage across different sports; in particular, it may help to differentiate this phenomenon between interdependent sports such as basketball or soccer (where teams members need more cooperation to complete tasks) and independent ones such as baseball.




%
%
%

\clearpage
\begin{adjustwidth}{-2.25in}{0in}
\section*{Supporting Information}

\subsection*{S1 Dataset}
\label{S1_dataset}
{\bf Dataset employed in this study.} Each line of the file corresponds to a game. The lines are formatted as follows: 
\\
\{\\
\{\{\text{time}, \text{score}\},...,\{\text{time}, \text{score}\}\},\\
\{\{\text{time}, \text{score}\},...,\{\text{time}, \text{score}\}\},\\
\{\{\text{time}, \text{score}\},...,\{\text{time}, \text{score}\}\},\\
\{\{\text{time}, \text{score}\},...,\{\text{time}, \text{score}\}\},\\
\}\\
\{\\
\{\{\text{time}, \text{score}\},...,\{\text{time}, \text{score}\}\},\\
\{\{\text{time}, \text{score}\},...,\{\text{time}, \text{score}\}\},\\
\{\{\text{time}, \text{score}\},...,\{\text{time}, \text{score}\}\},\\
\{\{\text{time}, \text{score}\},...,\{\text{time}, \text{score}\}\},\\
\}\\
\{\{``\text{team playing home}'', ``\text{team playing away}''\}, \\
\{\text{Final score of team playing home}, \text{Final score of team playing away}\}, \\
\{ ``\text{game year}'', ``\text{game month}''\}\}

\hspace{-0.5cm}The first set of brackets represent the evolution of the score for a team playing home and the second one is the same for a team playing away. The inner brackets correspond to each NBA quarter.%
\clearpage

\end{adjustwidth}

\begin{adjustwidth}{-2.25in}{0in}
\subsection*{S1 Fig}
\label{S1_Fig}
{\bf Average score $S(t)$ as function of the game time $t$ when playing at home (red circles) and away (blue squares).} Each panel shows the results for a NBA season (indicated in the plots) and the continuous lines (red for home and blue for away) represent the adjusted power-law models [$S(t) = R\, t^{\alpha}$]. 
\begin{center}
\includegraphics[scale=0.45]{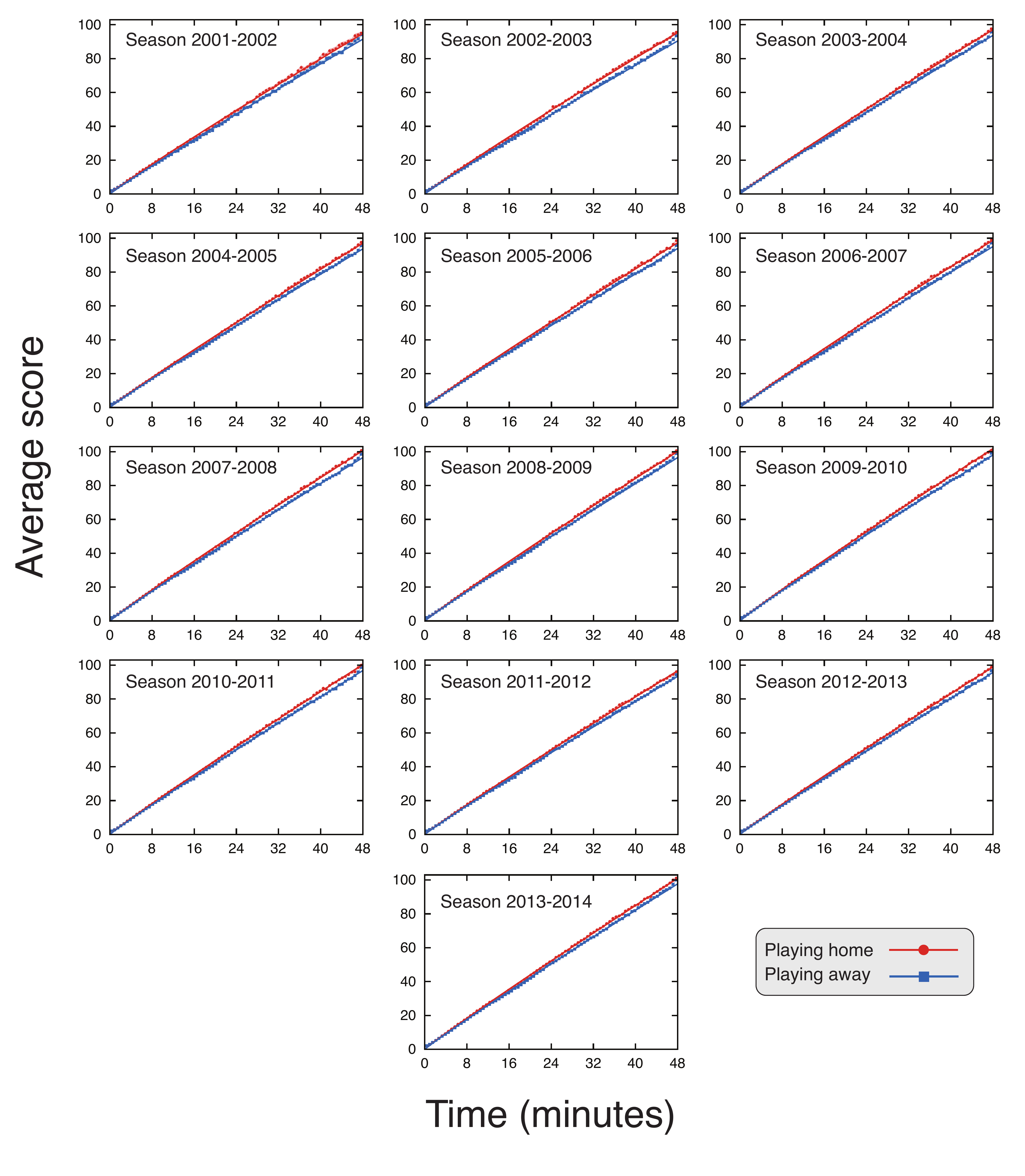}
\end{center}
\end{adjustwidth}
\clearpage

\begin{adjustwidth}{-2.25in}{0in}
\subsection*{S2 Fig}
\label{S2_Fig}
{\bf Average score $S(t)$ as function of the game time $t$ when playing home (red circles) and away (blue squares) for each NBA team along the seasons 2004--05 to 2013--14.} Each panel shows the results for a NBA team (indicated in the plots) and the continuous lines (red for home and blue for away) represent the adjusted power-law models [$S(t) = R\, t^{\alpha}$].
\begin{center}
\includegraphics[scale=0.27]{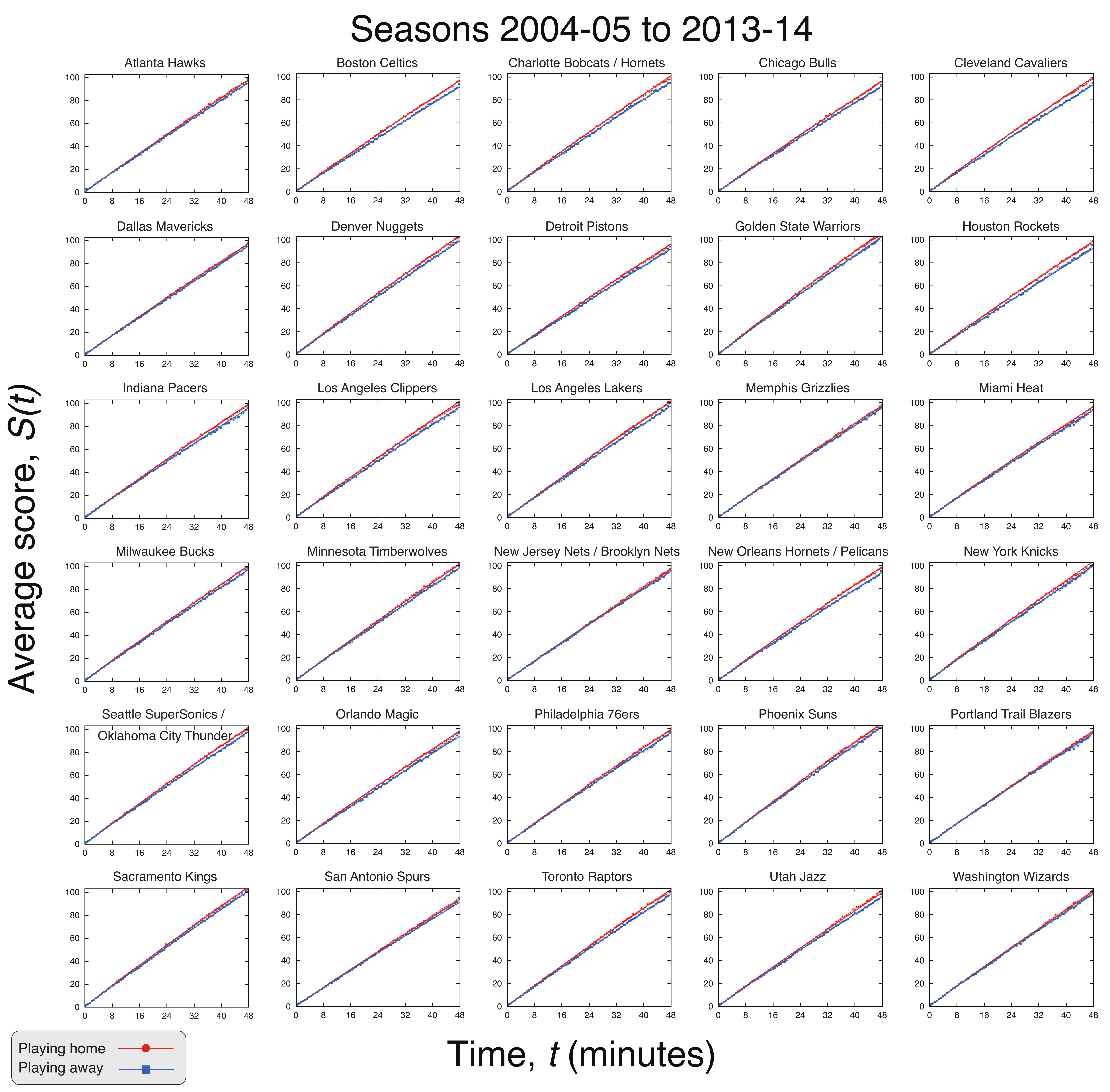}
\end{center}
\end{adjustwidth}
\clearpage

\begin{adjustwidth}{-2.25in}{0in}
\subsection*{S3 Fig}
\label{S3_Fig}
{\bf Cumulative distributions of the time intervals between stores when the teams play home (red dots) and away (blue squares) for the first quarter of the games.} The panels show the distributions in log-lin scale for each NBA season. The straight lines are guides for the eyes indicating the adjusted exponential behavior of these distributions.
\begin{center}
\includegraphics[scale=0.43]{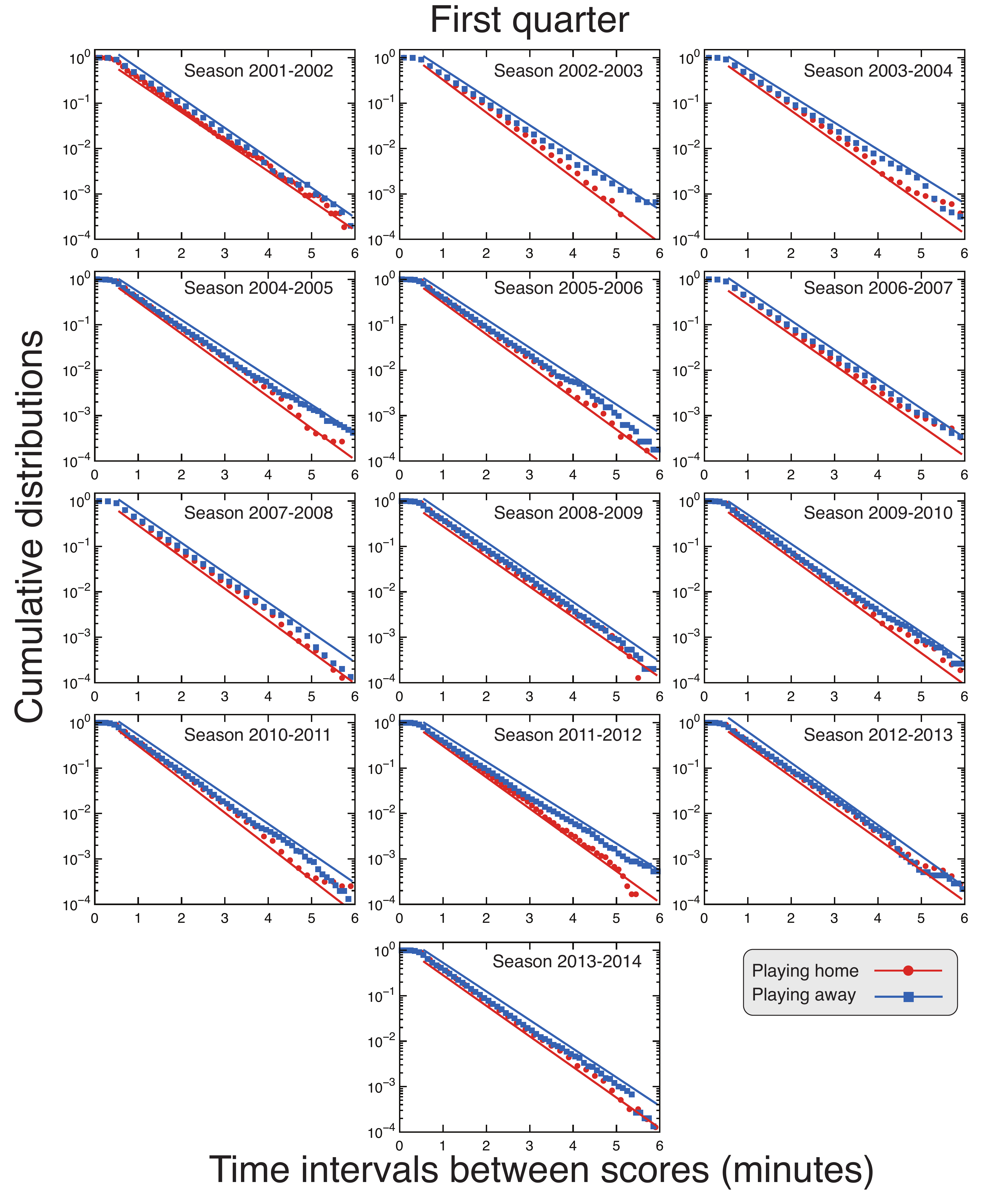}
\end{center}
\end{adjustwidth}
\clearpage

\begin{adjustwidth}{-2.25in}{0in}
\subsection*{S4 Fig}
\label{S4_Fig}
{\bf Cumulative distributions of the time intervals between stores when the teams play home (red dots) and away (blue squares) for the second quarter of the games.} The panels show the distributions in log-lin scale for each NBA season. The straight lines are guides for the eyes indicating the adjusted exponential behavior of these distributions.
\begin{center}
\includegraphics[scale=0.43]{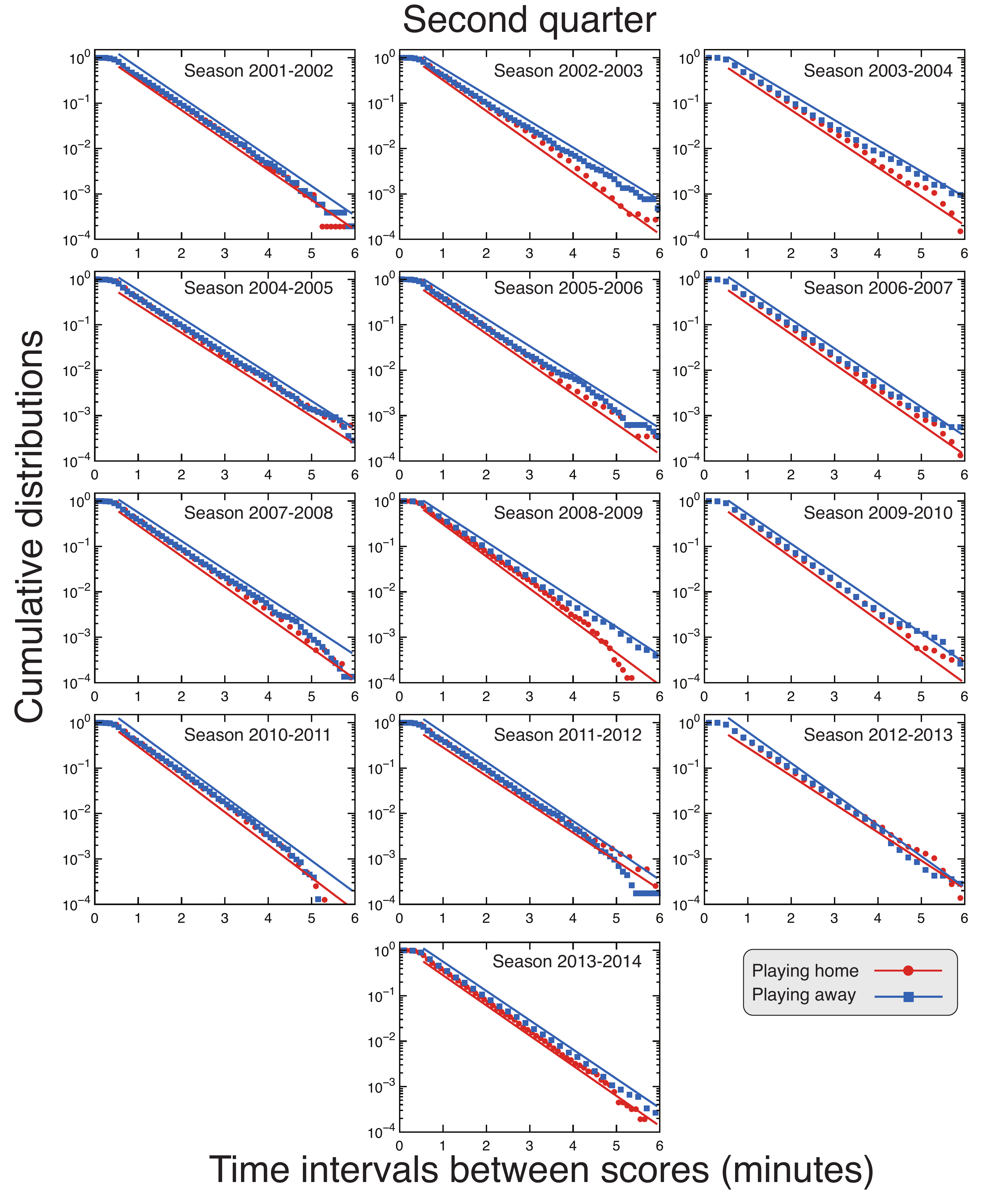}
\end{center}
\end{adjustwidth}
\clearpage

\begin{adjustwidth}{-2.25in}{0in}
\subsection*{S5 Fig}
\label{S5_Fig}
{\bf Cumulative distributions of the time intervals between stores when the teams play home (red dots) and away (blue squares) for the third quarter of the games.} The panels show the distributions in log-lin scale for each NBA season. The straight lines are guides for the eyes indicating the adjusted exponential behavior of these distributions.
\begin{center}
\includegraphics[scale=0.43]{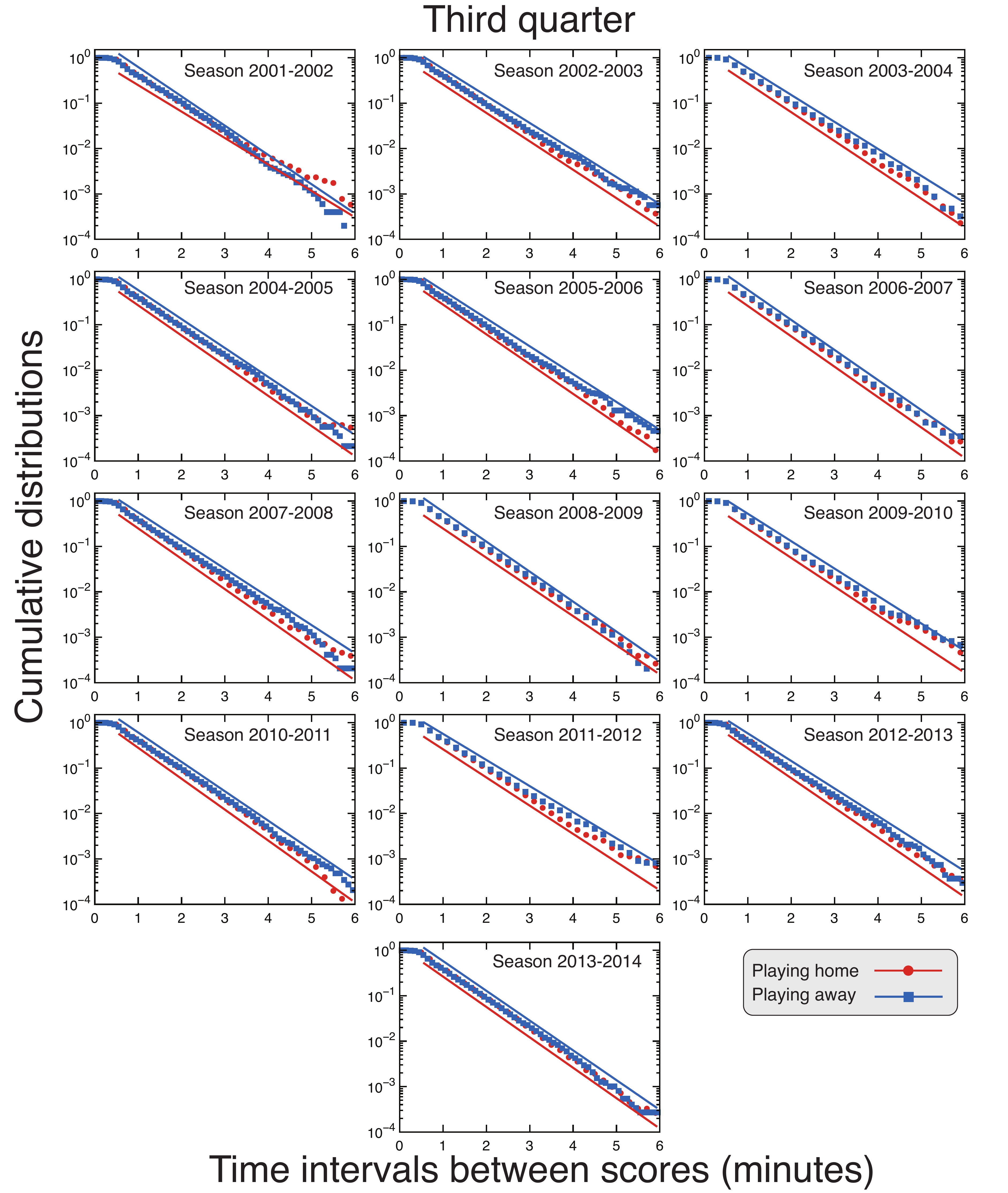}
\end{center}
\end{adjustwidth}
\clearpage

\begin{adjustwidth}{-2.25in}{0in}
\subsection*{S6 Fig}
\label{S6_Fig}
{\bf Cumulative distributions of the time intervals between stores when the teams play home (red dots) and away (blue squares) for the fourth quarter of the games.} The panels show the distributions in log-lin scale for each NBA season. The straight lines are guides for the eyes indicating the adjusted exponential behavior of these distributions.
\begin{center}
\includegraphics[scale=0.43]{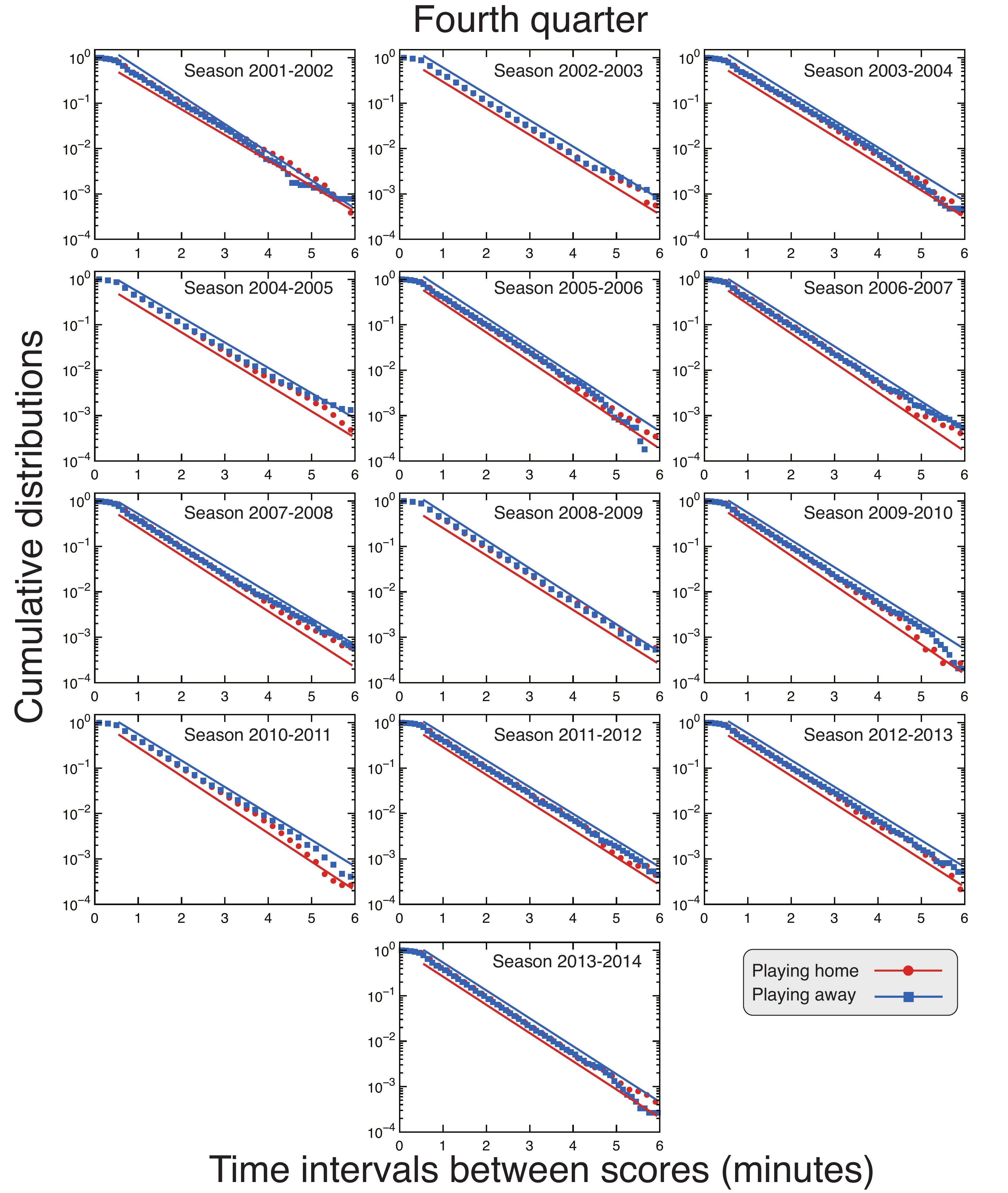}
\end{center}
\end{adjustwidth}
\clearpage

\begin{adjustwidth}{-2.25in}{0in}
\subsection*{S7 Fig}
\label{S7_Fig}\tabularnewline
{\bf Cumulative distributions of the time intervals between stores when the teams play home (red dots) and away (blue squares) for each NBA team along the seasons 2004--05 to 2013--14.} The dashed lines are guides for the eyes indicating the adjusted exponential behavior of these distributions.
\begin{center}
\includegraphics[scale=0.27]{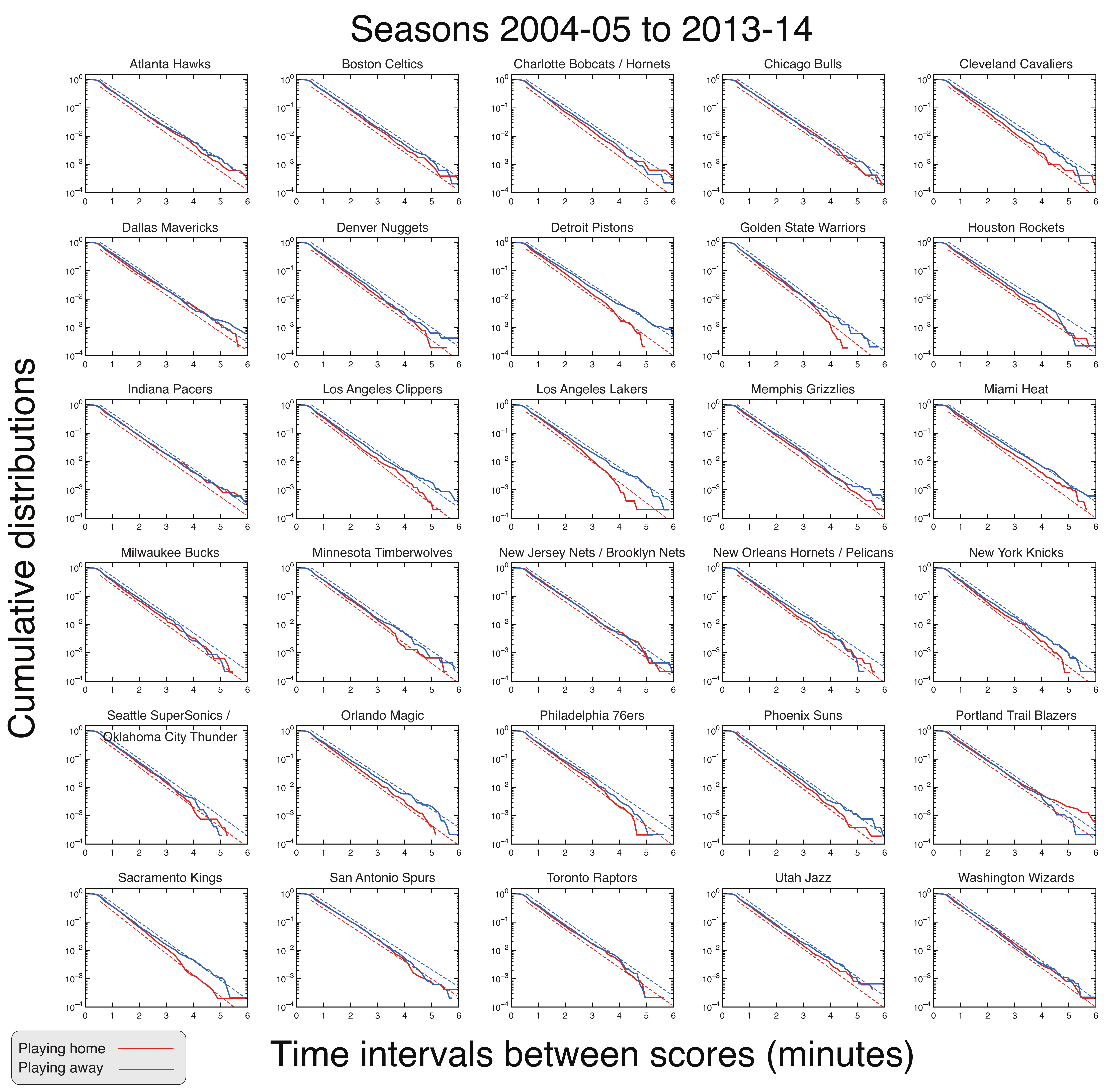}
\end{center}
\end{adjustwidth}
\clearpage

\begin{thebibliography}{10}
\bibitem{Kahn} Kahn L (2000) The sports business as a labor market laboratory. Journal of Economic Perspectives 14: 75-94. 
\bibitem{BenNaim} Ben-Naim E, Hengartner NW, Redner S, Vazquez F (2013) Randomness in Competitions. Journal of Statistical Physics 151: 458-474. doi:10.1007/s10955-012-0648-x
\bibitem{BenNaim2} Ben-Naim E, Redner S, Vazquez F (2007) Scaling in tournaments. Europhysics Letters 77: 30005. doi:10.1209/0295-5075/77/30005
\bibitem{BenNaim3} Ben-Naim E, Hengartner NW (2007) Efficiency of competitions. Physical Review E 76: 026106. doi:10.1103/PhysRevE.76.026106
\bibitem{Sire} Sire C (2007) Universal statistical properties of poker tournaments. Journal of Statistical Mechanics: Theory and Experiment 2007: P08013. doi:10.1088/1742-5468/2007/08/P08013
\bibitem{Sire2} Sire C, Redner S (2009) Understanding baseball team standings and streaks (2009) The European Physical Journal B 67: 473-481. doi:10.1140/epjb/e2008-00405-5
\bibitem{Heuer} Heuer A, Rubner O (2009) Fitness, chance, and myths: an objective view on soccer results. The European Physical Journal B 67: 445.
\bibitem{Ribeiro} Ribeiro HV, Mendes RS, Malacarne LC, Picoli Jr. S, Santoro PA (2010) Dynamics of tournaments: the soccer case. The European Physical Journal B 75: 327. doi:10.1140/epjb/e2010-00115-5
\bibitem{Gabel} Gabel A, Redner S (2012) Random Walk Picture of Basketball Scoring. Journal of Quantitative Analysis in Sports 8. doi:10.1515/1559-0410.1416
\bibitem{Ribeiro2} Ribeiro HV, Mukherjee S, Zeng XHT (2012) Anomalous diffusion and long-range correlations in the score evolution of the game of cricket. Physical Review E 86: 022102. doi:10.1103/PhysRevE.86.022102
\bibitem{Ribeiro3} Ribeiro HV, Mendes RS, Lenzi EK, Castillo-Mussot M, Amaral LAN (2013) Move-by-move dynamics of the advantage in chess matches reveals population-level learning of the game. PLoS ONE 8: e54165. doi:10.1371/journal.pone.0054165
\bibitem{Clauset} Clauset A, Kogan M, Redner S (2015) Safe leads and lead changes in competitive team sports. Physical Review E 91: 06281. doi:10.1103/PhysRevE.91.062815
\bibitem{BenNaim4} Ben-Naim E, Vazquez F, Redner S (2005) What is the most competitive sport?. arXiv:physics/0512143v1
\bibitem{DeSaa} de Sa\'a Guerra Y, Mart\'in Gonz\'alez JM, Arjonilla L\'opez N, Sarmiento Montesdeoca S, Rodr\'iguez Ruiz D, Garc\'ia-Manso JM (2011) Competitiveness analysis in the NBA regular seasons, Education Physical Training Sport 80: 17-21.
\bibitem{DeSaa3} de Sa\'a Guerra Y, Mart\'in Gonz\'alez JM, Sarmiento Montesdeoca S, Rodr\'iguez Ruiz D,  Garc\'ia-Rodr\'iguez A, Garc\'ia-Manso JM (2012) A model for competitiveness level analysis in sports competitions: Application to basketball, Physica A 391: 2997-3004. doi:10.1016/j.physa.2012.01.014
\bibitem{DeSaa2} de Sa\'a Guerra Y, Mart\'in Gonz\'alez JM, Sarmiento Montesdeoca S, Rodr\'iguez Ruiz D, Arjonilla L\'opez N, Garc\'ia-Manso JM (2013) Basketball scoring in NBA games: an example of complexity. Journal of Systems Science and Complexity 26: 94-103. doi:10.1007/s11424-013-2282-3
\bibitem{Erculj} Erčulj F, Štrumbelj E (2015) Basketball Shot Types and Shot Success in Different Levels of Competitive Basketball. PLoS ONE 10: e0128885. doi:10.1371/journal.pone.0128885
\bibitem{Duch} Duch J, Waitzman JS, Amaral LAN (2010) Quantifying the performance of individual players in a team activity. PLoS ONE 5: e10937. doi:10.1371/journal.pone.0010937
\bibitem{Radicchi} Radicchi F (2011) Who Is the Best Player Ever? A Complex Network Analysis of the History of Professional Tennis. PLoS ONE 6: e17249. doi:10.1371/journal.pone.0017249
\bibitem{Fewell} Fewell JH, Armbruster D, Ingraham J, Petersen A, Waters JS (2012) Basketball Teams as Strategic Networks. PLoS ONE 7: e47445. doi:10.1371/journal.pone.0047445
\bibitem{Mukherjee} Mukherjee S (2013) Complex Network Analysis in Cricket: Community structure, player's role and performance index. Advances in Complex Systems 16: 3. doi:10.1142/S0219525913500318
\bibitem{Mukherjee2} Mukherjee S (2013) Quantifying individual performance in Cricket - A network analysis of Batsmen and Bowlers. Physica A  393: 624-637. doi:10.1016/j.physa.2013.09.027
\bibitem{Sampaio} Sampaio J, McGarry T, Calleja-Gonz\'alez J, Jim\'enez S\'aiz S, Schelling i del Alcázar X, Balciunas M (2015) Exploring Game Performance in the National Basketball Association Using Player Tracking Data. PLoS ONE 10: e0132894. doi:10.1371/journal.pone.0132894
\bibitem{Radicchi2} Radicchi F (2012) Universality, Limits and Predictability of Gold-Medal Performances at the Olympic Games. PLoS ONE 7: e40335. doi:10.1371/journal.pone.0040335
\bibitem{Saavedra} Saavedra S, Mukherjee S, Bagrow JP (2012) Is coaching experience
associated with effective use of timeouts in basketball?. Scientific Reports 2: 676. doi:10.1038/srep00676.
\bibitem{Willer} Willer R, Sharkey A, Frey S (2012) Reciprocity on the Hardwood: Passing Patterns among Professional Basketball Players. PLoS ONE 7: e49807. doi:10.1371/journal.pone.0049807
\bibitem{Uhlmann} Uhlmann EL, Barnes CM (2014) Selfish Play Increases during High-Stakes NBA Games and Is Rewarded with More Lucrative Contracts. PLoS ONE 9: e95745. doi:10.1371/journal.pone.0095745
\bibitem{Petersen} Petersen AM, Jung WS, Yang JS, Stanley HE (2011) Quantitative and empirical demonstration of the Matthew effect in a study of career longevity. Proc. Natl. Acad. Sci. USA 108: 18-23. doi:10.1073/pnas.1016733108
\bibitem{Perc} Perc M (2014) The Matthew effect in empirical data. Journal of The Royal Society Interface 11: 20140378. doi:10.1098/rsif.2014.0378.
\bibitem{Yaari} Yaari G, Eisenmann S (2011) The Hot (Invisible?) Hand: Can Time Sequence Patterns of Success/Failure in Sports Be Modeled as Repeated Random Independent Trials?. PLoS ONE 6: e24532. doi:10.1371/journal.pone.0024532
\bibitem{Bock} Bock JR, Maewal A, Gough DA (2012) Hitting Is Contagious in Baseball: Evidence from Long Hitting Streaks. PLoS ONE 7: e51367. doi:10.1371/journal.pone.0051367
\bibitem{Yaari2} Yaari G, David G (2012) ``Hot Hand'' on Strike: Bowling Data Indicates Correlation to Recent Past Results, Not Causality. PLoS ONE 7: e30112. doi:10.1371/journal.pone.0030112
\bibitem{Csapo} Csapo P, Raab M (2014) ``Hand down, Man down.'' Analysis of Defensive Adjustments in Response to the Hot Hand in Basketball Using Novel Defense Metrics. PLoS ONE 9: e114184. doi:10.1371/journal.pone.0114184
\bibitem{Rosner} Rosner SR, Shropshire KL. The Business of Sports. 2th ed. Sudbury: Jones \& Bartlett Learning; 2010.
\bibitem{Schwartz} Schwartz B and Stephen F. Barsky SF (1977) The Home Advantage. Social Forces 55: 641-661.
\bibitem{Courneya} Courneya KS, Carron AV (1992) The home advantage in sport competitions: A literature review. Journal of Sport and Exercise Psychology: 14, 13-27.
\bibitem{Nevill2} Nevill AM, Holder LR (1999) Home Advantage in Sport: An Overview of Studies on the Advantage of Playing at Home. Sports Medicine 28: 221-236. doi:10.2165/00007256-199928040-00001
\bibitem{Nevill1} Nevill A, Balmer N, Wolfson S (2005) The extent and causes of home advantage: Some recent insights. Journal of Sports Sciences 23: 335-336. doi:10.1080/02640410500074375. 
\bibitem{Pollard} Pollard R (1986) Home advantage in soccer: A retrospective analysis, Journal of Sports Sciences 4: 237-248. doi:10.1080/02640418608732122
\bibitem{Nevill} Nevill AM, Newell SM, Gale S (1996) Factors associated with home advantage in English and Scottish soccer matches, Journal of Sports Sciences 14: 181-186. doi:10.1080/02640419608727700
\bibitem{Pollard3} Pollard R (2006) Worldwide regional variations in home advantage in association football. Journal of Sports Sciences 24: 231-240. doi:10.1080/02640410500141836
\bibitem{Page} Page L, Page K (2007) The second leg home advantage: Evidence from European football cup competitions. Journal of Sports Sciences 25: 1547-1556. doi:10.1080/02640410701275219
\bibitem{Monks} Monks J, Husch J (2009) The Impact of Seeding, Home Continent, and Hosting on FIFA World Cup Results. Journal of Sports Economics 10: 391-408. doi:10.1177/1527002508328757
\bibitem{Riedl} Riedl D, Strauss B, Heuer A, Rubner O (2014) Finale furioso: referee-biased injury times and their effects on home advantage in football. Journal of Sports Sciences 33: 327-336. doi:10.1080/02640414.2014.944558
\bibitem{Staufenbiel} Staufenbiel K, Lobinger B, Strauss B (2015) Home advantage in soccer
-- A matter of expectations, goal setting and tactical decisions of coaches?. Journal of Sports Sciences 33: 1932-1941. doi:10.1080/02640414.2015.1018929
\bibitem{Bray} Bray SR, Obara J, Kwan M (2005) Batting last as a home advantage factor in men's NCAA tournament baseball. Journal of Sports Sciences 23: 681-686. doi:10.1080/02640410400022136
\bibitem{Levernier} Levernier W, Barilla AG (2007) An Analysis of the Home-Field Advantage in Major League Baseball Using Logit Models: Evidence from the 2004 and 2005 Seasons. Journal of Quantitative Analysis in Sports 3: 1559-0410. doi:10.2202/1559-0410.1045
\bibitem{McGuire} McGuire EJ, Widmeyer WN, Courneya KS, Carron AV (1992) Aggression as a Potential Mediator of the Home Advantage in Professional Ice Hockey. Journal of Sport \& Exercise Psychology 14: 148-158.
\bibitem{Agnew} Agnew GA, Carron AV (1994) Crowd effects and the home advantage. International Journal of Sport Psychology 25: 53-62.
\bibitem{Gomezrh} G\'omez MA, Pollard R, Luis-Pascual JC (2011) Comparison of the home advantage in nine different professional team sports in Spain. Perceptual and motor skills 113: 150-156. doi:10.2466/05.PMS.113.4.150-156
\bibitem{Jones} Jones MB (2007) Home Advantage in the NBA as a Game-Long Process. Journal of Quantitative Analysis in Sports 3: 1559-0410. doi:10.2202/1559-0410.1081 
\bibitem{PollardB} Pollard R, G\'omez MA (2013) Variations in home advantage in the national basketball leagues of Europe. Revista de Psicolog\'ia del Deporte 22: 263-266.
\bibitem{Thomas} Thomas S, Reeves C, Bellhome A (2008) Home advantage in the Six Nations Rugby Union Tournament. Perceptual and Motor Skills 106: 113-116. doi:10.2466/PMS.106.1.113-116
\bibitem{Clarkeaf} Clarke SR (2005) Home advantage in the Australian football league. Journal of Sports Sciences 23: 375-385. doi:10.1080/02640410500074391
\bibitem{Prietowp} Prieto J, G\'omez MA, Pollard R (2013) Home Advantage in Men's and Women's Spanish First and Second Division Water Polo Leagues. Journal of human kinetics 37: 137-143. doi:10.2478/hukin-2013-0034
\bibitem{Marcelino} Marcelino R, Mesquita I, Palao JM, Sampaio J (2009) Home advantage in high-level volleyball varies according to set number. Journal of Sports Science and Medicine 8: 352-356.
\bibitem{Pollardhb} Pollard R, G\'omez MA (2012) Re-assessment of home advantage in Spanish handball: comment on Gutierrez, et al. (2012). Perceptual \& Motor Skills 115: 937-943. doi:10.2466/06.05.PMS.115.6.937-943
\bibitem{Oliveira} Oliveira T, G\'omez MA, Sampaio J (2012) Effects of game location, period, and quality of opposition in elite handball performances, Perceptual and Motor Skills 114: 783-794. doi:10.2466/30.06.PMS.114.3.783-794
\bibitem{Smiatek} Smiatek J, Heuer A (2012) A statistical view on team handball results: home advantage, team fitness and prediction of match outcomes. arXiv:1207.0700
\bibitem{Morley} Morley B, Thomas D (2005) An investigation of home advantage and other factors affecting outcomes in English one-day cricket matches. Journal of Sports Sciences 23: 261-268. doi:10.1080/02640410410001730133
\bibitem{Koning} Koning RH (2011) Home advantage in professional tennis. Journal of Sports Sciences 29: 19-27. doi:10.1080/02640414.2010.516762
\bibitem{Nevill4} Nevill AM, Holder RL, Bardsley A, Carvert H, Jones S (1997) Identifying home advantage in international tennis and golf tournaments. Journal of Sports Sciences 15: 437-443. doi:10.1080/026404197367227
\bibitem{Balmer} Balmer NJ, Nevill AM, Williams AM (2001) Home advantage in the Winter Olympics (1908-1998). Journal of Sports Sciences 19: 129-139. doi:10.1080/026404101300036334
\bibitem{Balmerso} Balmer NJ, Nevill AM, Williams AM (2003) Modelling home advantage in the Summer Olympic Games. Journal of Sports Sciences 21: 469-478. doi:10.1080/0264041031000101890
\bibitem{Jones2} Jones MB (2013) The home advantage in individual sports: An augmented review. Psychology of Sport and Exercise 14: 397-404. doi:10.1016/j.psychsport.2013.01.002
\bibitem{Jamieson} Jamieson JP (2010) The home field advantage in athletics: A meta-analysis. Journal of Applied Social Psychology 40: 1819-1848. doi:10.1111/j.1559-1816.2010.00641.x
\bibitem{Nevill3} Nevill AM, Balmer NJ, Williams AM (2002) The influence of crowd noise and experience upon refereeing decisions in football. Psychology of Sport and Exercise 3: 261-272. doi:10.1016/S1469-0292(01)00033-4
\bibitem{Myers} Myers TD (2014) Achieving external validity in home advantage research: generalizing crowd noise effects. Frontiers in Psychology 5: 532. doi:10.3389/fpsyg.2014.00532
\bibitem{Anders} Anders A, Rotthoff KW (2014) Is home-field advantage driven by the fans? Evidence from across the ocean. Applied Economics Letters 21: 1165-1168. doi:10.1080/13504851.2014.914139
\bibitem{Goumas} Goumas C (2014) Home advantage in Australian soccer. Journal of Science and Medicine in Sport 17: 119-123. doi:10.1016/j.jsams.2013.02.014
\bibitem{Gomezta} G\'omez MA, Lorenzo A, Ib\'a\~nez SJ, Ortega E, Leite N, Sampaio J (2010). An analysis of defensive strategies used by home and away basketball teams. Perceptual and Motor Skills 110: 159-166. doi:10.2466/pms.110.1.159-166
\bibitem{Pollardta} Pollard R (2008) Home advantage in football: A current review of an unsolved puzzle. The Open Sports Sciences Journal 1: 12-14. doi:10.2174/1875399X00801010012
\bibitem{Pollard2} Pollard R (2002) Evidence of a reduced home advantage when a team moves to a new stadium. Journal of Sports Sciences 20: 969-973. doi:10.1080/026404102321011724
\bibitem{Johnston} Johnston R (2008) On referee bias, crowd size, and home advantage in the English soccer Premiership, Journal of Sports Sciences 26: 563-568. doi:10.1080/02640410701736780
\bibitem{Gomes} G\'omez MA, Pollard R (2011) Reduced home advantage for basketball teams from capital cities in Europe. European Journal of Sport Science 11: 143-148. doi:10.1080/17461391.2010.499970
\bibitem{Pollard5} Pollard R, Pollard G (2005) Long-term trends in home advantage in professional team sports in North America and England (1876-2003). Journal of Sports Sciences 23: 337-350. doi:10.1080/02640410400021559
\bibitem{Pollard6} Pollard R (2006) Worldwide regional variations in home advantage in association football, Journal of Sports Sciences 24: 231-240. doi:10.1080/02640410500141836
\bibitem{Pollard4} Pollard R, G\'omez MA (2014) Comparison of home advantage in men's and women's football leagues in Europe, European Journal of Sport Science 14: S77-S83. doi:10.1080/17461391.2011.651490
\bibitem{Sheskin} Sheskin DJ. The Handbook of Parametric and Nonparametric Statistical Procedures. 5th ed. New York: Chapman \& Hall/CRC Press; 2011.

\end{thebibliography}
\end{document}